%% file: TemporalBell.tex
\documentclass[a4paper,11pt]{article}
\pdfoutput=1 
\usepackage{jcappub}
\usepackage{amsmath}
\usepackage{graphicx}
\usepackage{latexsym}
\usepackage{xspace}
\usepackage{color}
\usepackage{hyperref} 
\usepackage{bm}
\usepackage{relsize}
\usepackage{tabularx}
\usepackage{multirow}
\usepackage{amssymb}
\usepackage[table]{xcolor}
\usepackage{braket}
\usepackage{comment}
\usepackage[utf8]{inputenc}
\usepackage{slashed}

\makeatletter
\gdef\@fpheader{}
\g@addto@macro\bfseries{\boldmath}
\makeatother

\newcommand{\prn}[1]{\left( {#1} \right)}
\newcommand{\com}[1]{\left[ {#1} \right]}

\newcommand{\bvec}[1]{\bm{#1}}
\newcommand{\abs}[1]{\left\vert {#1} \right\vert}

\newcommand{\RR}{{\mathrm R}}
\newcommand{\II}{{\mathrm I}}
\newcommand{\erfc}{{\mathrm{erfc}}}

\input{newcommands}

\subheader{}

\title{Bipartite temporal Bell inequalities \\for two-mode squeezed states}

\author[a]{Kenta Ando,}
\affiliation[a]{ICRR, University of Tokyo, Kashiwa, 277-8582, Japan}
\emailAdd{ando@icrr.u-tokyo.ac.jp}

\author[b]{Vincent Vennin}
\affiliation[b]{Laboratoire Astroparticule et Cosmologie, CNRS, Universit\'e de Paris, 75013 Paris, France}
\emailAdd{vincent.vennin@apc.univ-paris7.fr}

\date{today}

\begin{document}
\sloppy

\abstract{
Bipartite temporal Bell inequalities are similar to the usual Bell inequalities except that, instead of changing the direction of the polariser at each measurement, one changes the time at which the measurement is being performed. By doing so, one is able to test for realism and locality, but relying on position measurements only. This is particularly useful in experimental setups where the momentum direction cannot be probed (such as in cosmology for instance). We study these bipartite temporal Bell inequalities for continuous systems placed in two-mode squeezed states, and find some regions in parameter space where they are indeed violated. We highlight the role played by the rotation angle, which is one of the three parameters characterising a two-mode squeezed state (the other two being the squeezing amplitude and the squeezing angle). In single-time measurements, it only determines the overall phase of the wavefunction and can therefore be discarded, but in multiple-time measurements, its time dynamics becomes relevant and crucially determines when bipartite temporal Bell inequalities can be violated. Our study opens up the possibility of new experimental designs for the observation of Bell inequality violations.}



\maketitle


\section{Introduction}
\label{sec:intro}
Quantum theory allows for the existence of correlations that display counter-intuitive properties, which are in stark contrast with our everyday experience of the macroscopic world. One such well-known example is Bell inequalities~\cite{Bell:1964kc}. In the Clauser-Horne-Shimony-Holt (CHSH) scenario~\cite{Clauser:1969ny}, two observers, commonly dubbed Alice and Bob, perform measurements of dichotomic variables $\hat{S}^a$ (for instance, spin variables, where ``$a$'' labels the direction of the polariser) for two subsystems $1$ and $2$ at separate spatial locations $x_1$ and $x_2$, and build the correlator $E(a,b)=\langle \hat{S}_1^a \hat{S}_2^b \rangle$. Under the assumptions of \emph{realism} and \emph{locality}, the following inequality holds,
\bea
\label{eq:Bell:inequality}
B=E(a,b)+E(a,b')+E(a',b)-E(a',b')\leq 2\, .
\eea 
In this context, ``realism'' is the assumption that, at each instant in time, the system definitely lies in one of several distinct configurations. The state of the system determines all measurement outcomes exactly, such that all observables possess pre-existing definite values. The principle of locality states that a system cannot be influenced by a spacelike-separated event, and rules out the concept of instantaneous ``action at a distance''.

Other types of Bell inequalities exist, which relie on performing measurements at different times, instead of at different locations. In the temporal Bell inequalities~\cite{1993PhRvL..71.3235P, 2004quant.ph..2127B, 2010NJPh...12h3055F}, Alice performs a projective measurement of $\hat{S}^a$ at time $t_1$ (so that the state collapses to an eigenstate of $\hat{S}^a$ upon the measurement), then Bob measures $\hat{S}^b$, on the same system, at a latter time $t_2$. They can then construct the correlators $E(a,b) = \langle \lbrace \hat{S}^a(t_1) , \hat{S}^b(t_2)\rbrace \rangle/2$, which satisfy the inequality~\eqref{eq:Bell:inequality}. Two main differences with the spatial Bell inequalities should be highlighted. First, in the temporal Bell inequalities, there is no need to have available a bipartite system made of two entangled sub-systems at two different locations, since repeated measurements are performed on the same system. Second, temporal Bell inequalities do not rest on the same assumptions. Although realism is still necessary, locality is now replaced with \emph{non-invasiveness}, \ie the ability to perform a measurement without disturbing the state of the system. One's everyday experience of the macroscopic world is that it is both realist and subject to non-invasive measurements, while in quantum mechanics, quantum superposition violates realism and the reduction of the wavefunction (in the Copenhagen interpretation) violates non-invasiveness. This explains why the inequality~\eqref{eq:Bell:inequality} can be violated by quantum systems.

Although the assumption of non-invasiveness plays the exact same role for temporally separated systems as locality does for spatially separated systems (spatial scenarios test local hidden variable theories while temporal scenarios test non-invasive hidden variable theories), these two types of experiments come with different loopholes. For temporal Bell experiments, the \emph{clumsiness loophole}~\cite{2012FoPh...42..256W} is the fundamental impossibility to prove that a physical measurement is actually non-invasive. It is the analogue of the \emph{communication loophole}~\cite{1978RPPh...41.1881C} in spatial Bell experiments. However, the communication loophole can be closed by making sure that two measurements are space-like separated, and special relativity ensures that events at one detector cannot influence measurements performed by the second detector. Such a solution to the clumsiness loophole does not exist.\footnote{Let us note that the \emph{freedom of choice} loophole, \ie the ability to ensure that the choice of measurement settings is ``free and random'', and independent of any physical process that could affect the measurement outcomes, can be closed, or at least pushed back to billions of years ago, by using cosmological sources, see \Refa{Rauch:2018rvx}.}

Testing non-invasiveness is still interesting since in alternatives to the Copenhagen interpretation of quantum mechanics, such as in dynamical collapse theories~\cite{Ghirardi:1985mt,Pearle:1988uh,Ghirardi:1989cn,Bassi:2003gd,Martin:2019jye}, the dynamics of projective measurements is altered, and investigating temporal correlations is thus likely to provide ways to distinguish between alternative ``interpretations''. Multiple-time measurements also enlarge the class of systems, and experimental setups, in which Bell inequality violations can be performed.  

For instance, if measurements are now performed at three (or more) different times, it is possible to measure a single observable $\hat{S}$ (\ie to fix the label $a$ in the above notations), and upon defining the two-time correlator $E(t_1,t_2)=\langle \lbrace \hat{S}(t_1), \hat{S}(t_2) \rbrace \rangle/2$, realism and non-invasiveness imply the inequality $-3\leq E(t_1,t_2)+E(t_2,t_3)-E(t_1,t_3)\leq 1$. This is the so-called Leggett-Garg inequality~\cite{Leggett:1985zz} (for a review, see \Refa{2014RPPh...77a6001E}), which can be generalised to higher-order strings of multiple-time measurements. In contrast to spatial Bell experiments, the Legget-Garg setup does not require to measure several non-commuting observables (\ie spins in different directions for instance), since a single observable does not commute with itself at different times in general so this is enough to ensure a non-commuting algebra to be present in the problem. This is particularly useful in contexts where only one spin operator can be measured, as is the case in cosmology~\cite{Martin:2015qta, Martin:2016nrr, Martin:2017zxs, Kanno:2017dci, Kanno:2017teu}: there, density perturbations propagate a growing mode, which plays the role of ``position'', and a decaying mode, which plays the role of ``momentum''. In practice, the decaying mode is too small to be detected by measurements of the large-scale structure of the universe, hence only the growing mode can be probed, which prevents one from performing experiments involving several, non-commuting spin operators. 

Finally, a last class of experiments exists, which mixes features of the two previous classes, and which is in fact what was originally considered by John Bell in 1966~\cite{1986NYASA.480..263B} (see \Refa{Martin:2019wta} for a recent and insightful resurrection of this work). There, one performs measurements separated both in space and time. More precisely, given a bipartite system made of two subsystems $1$ and $2$, located at two different locations $x_1$ and $x_2$, Alice measures the same dichotomic variable $\hat{S}_1$ on the first sub-system at times $t_a$ and $t_{a}'$, while Bob measures $\hat{S}_2$ on the second sub-system at times $t_b$ and $t_{b}'$. Here, the two sets of measuring events, $\lbrace (x_1,t_a), (x_1,t_{a}')\rbrace$ on one hand, and $\lbrace (x_2,t_b), (x_2,t_{b}')\rbrace$ on the other hand, are causally disconnected, and time plays the exact same role as the measurement parameter (\eg the polariser angle) in the ordinary spatial Bell inequalities. We call this kind of setup ``bipartite temporal Bell inequality''. The inequality~\eqref{eq:Bell:inequality} is satisfied provided the assumptions of realism and locality hold (if Alice and/or Bob perform repeated measurements on the same physical realisation of the system, or if the two sets of measuring events are not causally disconnected, then one must add non-invasiveness, but this is not compulsory), so the same fundamental properties are tested as in the usual spatial Bell inequality. However, compared to the spatial Bell inequality, this has the advantage of relying on measuring a single spin operator. 
%
\begin{table*}[t]
\newcolumntype{L}[1]{>{\raggedright\let\newline\\\arraybackslash\hspace{0pt}}m{#1}}
\newcolumntype{C}[1]{>{\centering}m{#1}}
\newcolumntype{R}[1]{>{\raggedleft\let\newline\\\arraybackslash\hspace{0pt}}m{#1}}
\begin{tabular}{{|C{4.0cm}||C{3.3cm}|C{3.1cm}|C{3.2cm}|}}
\hline
Type of inequality &  Assumptions & Requires bipartite system & involves single spin measurement only \tabularnewline \hline \hline
Spatial Bell & realism and locality & yes & no
\tabularnewline \hline
Temporal Bell & realism and non-invasiveness & no & no
\tabularnewline \hline
Legget-Garg &  realism and non-invasiveness & no & yes
\tabularnewline \hline
Bipartite temporal Bell & realism and locality & yes & yes
\tabularnewline \hline
\end{tabular}
\caption[]{Classes of Bell inequality experiments discussed in \Sec{sec:intro}.} 
\label{tab:summary} 
\end{table*}
%

In table~\ref{tab:summary}, we summarise the main features of the four classes of Bell inequalities discussed above: spatial Bell inequalities, temporal Bell inequalities, Legget-Garg inequalities and bipartite temporal Bell inequalities. In this work we focus on bipartite temporal Bell inequalities, since they are the only ones that allow us to test realism and locality, while relying on measurements of a single spin operator. In practice, we consider continuous-variable systems that are placed in a two-mode squeezed state~\cite{Caves:1985zz,Schumaker:1985zz}. These states are entangled states that arise in a large variety of physical situations, since any quadratic Hamiltonian produces squeezed states. They are therefore commonly found in quantum optics~\cite{1987PhRvL..59.2555H, 2005PhRvL..95x3603V, 2015PhRvP...3d4005D}.\footnote{For a quantum field on a statistically isotropic background, the Fourier modes corresponding to opposite wavenumbers $\vec{k}$ and $-\vec{k}$ also evolve towards a two-mode squeezed states. In particular, this is the case of primordial cosmological perturbations~\cite{Starobinsky:1980te,Guth:1980zm,Linde:1981mu}. However, in that case, the two subsystems, ``$\vec{k}$'' and ``$-\vec{k}$'', correspond to disconnected regions in Fourier space, not in real space, so ``locality'' would then be tested in Fourier space, which may not be as relevant.} In the large-squeezing limit, they also provide a realisation of the Einstein-Podolsky-Rosen (EPR) state~\cite{Einstein:1935rr}. Let us note that the spatial Bell inequalities, and the Legget-Garg inequalities, have been already applied to two-mode squeezed states in \Refa{Martin:2016tbd} and in \Refa{Martin:2016nrr} respectively, and in this work we perform a similar analysis for the bipartite temporal Bell inequalities.

This paper is organised as follows. In \Sec{sec:setup}, we introduce a pseudo-spin operator for continuous systems and show how its projective, two-time correlator can be computed in a two-mode squeezed state. In \Sec{sec:Analytical:Limits}, we investigate several limiting cases, in order to gain analytical insight into what would be otherwise a tedious, high-dimensional parameter space to explore. In \Sec{sec:Numerical:Exploration}, we present numerical results, and show that bipartite temporal Bell inequalities can indeed be violated by two-mode squeezed states. We present our conclusions in \Sec{sec:conslusion}, and we end the paper by four appendices, to which we defer several technical aspects of our calculation.
\section{Two-time correlators}
\label{sec:setup}
%
\subsection{Projective measurements}
The investigation of temporal Bell inequalities requires to define quantum expectation values of projective measurements. Indeed, in spatial Bell inequalities, the two operators $\hat{S}_1^a$ and $\hat{S}_2^b$ commute (since they act on two separate subsystems), hence $\hat{S}_1^a \hat{S}_2^b$ is Hermitian and the correlator is simply given by $E(a,b)=\langle \hat{S}_1^a \hat{S}_2^b \rangle$. However, in temporal Bell inequalities, one has to work in the Heisenberg picture and consider the time-evolved operators
\bea
\hat{S}_1(t_a) = \hat{U}^\dagger(t_a) \hat{S}_1 \hat{U}(t_a)
\quad\quad\mathrm{and}\quad\quad
\hat{S}_2(t_b) = \hat{U}^\dagger(t_b) \hat{S}_2 \hat{U}(t_b)
\eea
where $\hat{U}(t)$ is the unitary time evolution operator. Although $\hat{S}_1$ and $\hat{S}_2$ commute, it is not the case in general for $\hat{S}_1(t_a)$ and $\hat{S}_2(t_b)$, hence $\hat{S}_1(t_a)\hat{S}_2(t_b)$ is not Hermitian and taking its expectation value would not give a real result. One must instead define quantum expectation values of projective measurements, which is done here following the prescription of \Refa{2010NJPh...12h3055F}. For explicitness, let us assume that $t_a\leq t_b$ although we will see that our final result also applies to the opposite situation. For the dichotomic variable $\hat{S}_1(t_a)$, with possible outcomes $\pm 1$, the projection operators onto the $+1$-eigenspace and the $-1$-eigenspace are respectively given by 
\bea
\hat{P}_{1}^{(1)}=\frac{1}{2}\left[1+\hat{S}_1\left(t_a\right)\right]
\quad\quad\mathrm{and}\quad\quad
\hat{P}_{-1}^{(1)}=\frac{1}{2}\left[1-\hat{S}_1\left(t_a\right)\right],
\eea
and similar expressions for $\hat{P}_{1}^{(2)}$ and $\hat{P}_{-1}^{(2)}$, the projection operators onto the $+1$-eigenspace and the $-1$-eigenspace of $\hat{S}_1(t_b)$. Let us denote by $r$ and $s$ the outcomes of the first and second measurements respectively. We denote by $P(r,s)$ the joint probability for Alice to get the outcome $r$ and for Bob to get the outcome $s$, which can be expressed as the probability that Alice observes the outcome $r$, multiplied by the probability that Bob gets the outcome $s$ upon measuring the state of the system after it has collapsed due to Alice's measurement. According to the projection postulate, after performing the measurement of $\hat{S}_1(t_a)$ on the state $\Ket{\psi}$, if the outcome of the measurement is $r=+1$, the state becomes $\hat{P}_{1}^{(1)}\Ket{\psi}$, while if the outcome of the measurement is $r=-1$, the state becomes $\hat{P}_{-1}^{(1)}\Ket{\psi}$. After the first measurement, $\Ket{\psi}$ thus becomes $\hat{P}_{r}^{(1)}\Ket{\psi}$, so according to the Born rule, the joint probability is given by
\bea
P(r,s) &=& \Braket{\psi | \hat{P}_{r}^{(1)} \hat{P}_{s}^{(2)} \hat{P}_{r}^{(1)} | \psi} \\
&=&\frac{1}{4}
+\frac{r}{4} \Braket{\psi | \hat{S}_1\left(t_a\right) | \psi}
+\frac{s}{8} \Braket{\psi | \hat{S}_2\left(t_b\right) | \psi}
+\frac{rs}{8} \Braket{\psi | \left\lbrace \hat{S}_1\left(t_a\right) , \hat{S}_2\left(t_b\right) \right\rbrace | \psi}
\nonumber \\ & & 
+\frac{s}{8} \Braket{\psi | \hat{S}_1\left(t_a\right) \hat{S}_2\left(t_b\right) \hat{S}_1\left(t_a\right) | \psi} \, ,
\label{eq:joint:proba}
\eea
where we have used that, since $\hat{S}_1\left(t_a\right)$ is a spin operator, $\hat{S}^2_1\left(t_a\right)=1$, and that $r^2=1$, and where $\lbrace\cdot,\cdot\rbrace$ denotes the anti-commutator. We now introduce the correlator
\bea
E\left(t_a,t_b\right) = \sum_{r,s} r s P(r,s) = P(+1,+1)-P(+1,-1)-P(-1,+1)+P(-1,-1)\, .
\eea
Plugging \Eq{eq:joint:proba} into that formula, one obtains
\bea
\label{eq:correlator:def}
E\left(t_a,t_b\right) = \frac{1}{2} \left\langle \psi \right. \left\vert \left\lbrace \hat{S}_1\left(t_a\right) , \hat{S}_2\left(t_b\right) \right\rbrace \right\vert \left.\psi\right\rangle\, .
\eea
One can check that, by construction, this correlator is real since taking the anticommutator guarantees that the operator $ \left\lbrace \hat{S}_1\left(t_a\right)  ,\hat{S}_2\left(t_b\right) \right\rbrace$ is Hermitian. Let us also note that although we assumed $t_a\leq t_b$, the result is perfectly symmetric in $t_a$ and $t_b$ so the correlator does not depend on who between Alice and Bob measures first. 
\subsection{Two-mode squeezed states}
The goal is to compute these correlators, and to test for the validity of the inequality~\eqref{eq:Bell:inequality}. In practice, this is done for two-mode squeezed states which we now introduce. A two-mode squeezed state can be obtained by evolving the vacuum state under a quadratic Hamiltonian; for instance, the Hamiltonian of a harmonic oscillator with a time-dependent frequency (see \Refa{Grain:2019vnq} for a recent pedagogical review of the squeezing formalism). The time evolution is described by a unitary operator $\hat U(t) = \hat U_{\mathrm{S}}(t)\hat R(t)$ composed of the squeezing operator
\bea
\label{eq:US:def}
	\hat U_{\mathrm{S}}(t) = \exp\left(re^{-2i\varphi}\hat c^\dagger_1\hat c^\dagger_2 - re^{2i\varphi}\hat c_1\hat c_2  \right)\, ,
\eea
and the rotation operator
\bea
\label{eq:R:def}
	\hat R(t) &=& \exp\left(i\theta \hat c_1^\dagger\hat c_1 +i\theta \hat c_2^\dagger\hat c_2\right)= e^{i\theta \hat n_1}e^{i\theta \hat n_2} \, .
\eea
Here, $\hat c_i$ and $\hat c_i^\dagger$ denote the annihilation operator and the creation operator in the subsystem $i=1,2$. They satisfy the commutation relations $\com{\hat c_i,\hat c_j^\dagger}=\delta_{ij}$, and $\hat{n}_i=\hat{c}_i^\dagger \hat{c}_i$ is the number of particles operator. The two-mode squeezed state is obtained as $\Ket{\Psi_{\mathrm{2sq}}(t)} = \hat U(t) \Ket{0,0}$, and in \App{App:ket:tmss}, we show that it can be written as
\bea
 \label{2sq}
	\Ket{\Psi_{\mathrm{2sq}}(t)} 
	&=& \frac{1}{\cosh (r)} \sum_{n = 0}^\infty e^{-2 i n \varphi} \tanh^n (r) \Ket{n,n}
\eea
in the basis of the number of particles ($\Ket{n,m}$ denotes the state with $n$ particles in the sector $1$ and $m$ particules in the sector $2$), see \Eq{eq:tmss:n:app}. 

These states are characterised by three time-dependent parameters: the squeezing amplitude $r$, the squeezing angle $\varphi$ and the rotation angle $\theta$.
Note that the expression for a two-mode squeezed state, \Eq{2sq}, does not include the rotation angle $\theta$ since the vacuum state is invariant under rotations, $\hat R(t)\Ket{0,0} =\Ket{0,0}$. This implies that measurements performed at the same time are insensitive to $\theta$, since $\theta$ simply adds an overall phase to the wavefunction. However, as will be made explicit below, when multiple-time measurements are performed, this is not the case anymore and the result becomes sensitive to the change in the overall phase between the measurement times. This can be interpreted as a consequence of the fact that, after performing a first measurement, the two-mode squeezed state is projected onto the eigenstate of a spin operator, which is not a two-mode squeezed state anymore, and which is therefore not invariant under rotations. This is why, contrary to what is usually done, $\theta$ is carefully kept in the calculations hereafter. 
\subsection{Spin operators for continuous variables}
Two-mode squeezed states describe bipartite continuous systems, and the investigation of Bell inequalities first requires to build dichotomic observables for such systems. Following \Refa{2004PhRvA..70b2102L}, we thus introduce
\bea
\label{eq:Si:def}
	\hat S_i(\ell) = \sum_{n = -\infty}^\infty(-1)^n\int_{n\ell}^{(n+1)\ell}\dd Q_i\Ket{Q_i}\Bra{Q_i}
\eea
where $\Ket{Q}_i$ is an eigenstate of the position operator for the subsystem $i$,
\bea
	\hat Q_i &=& \frac{1}{\sqrt{2}}\prn{\hat c_i + \hat c_i^\dagger} .
	\label{q_def}
\eea
In practice, measuring $S_i$ can be done by measuring $Q_i$, identifying in which interval $[n\ell, (n+1)\ell)$ of size $\ell$ it lies, and returning $(-1)^n$. The variable $S_i$ is therefore dichotomic (it can take values $+1$ or $-1$), as can also be explicitly checked from \Eq{eq:Si:def} by verifying that $\hat{S}_i^2=1$. In the limit where $\ell$ is infinite, $\hat{S}_i$ reduces to the sign operator,
\bea
\label{eq:Spin:Large:L:sign}
    \hat S_i(\ell\rightarrow\infty) = \mathrm{sign}\left({\hat Q_i} \right).
\eea
As explained in \Refa{2004PhRvA..70b2102L}, one could define two other operators $\hat{S}_x(\ell)$, $\hat{S}_y(\ell)$, such that together with $\hat{S}$ they obey the standard $SU(2)$ commutation relations, making $\hat S$ an actual pseudo-spin operator. Here, we will only need $\hat{S}$. Since its determination rests on position measurements, the Bell experiment we propose is therefore based on position measurement only, which as we explained above is convenient for situations in which the conjugated momentum cannot be accessed. 

Let us also mention that there are other ways to define dichotomic variables for continuous systems, see \Refs{PhysRevLett.82.2009, PhysRevLett.88.040406, 2016PhRvA..94a2105L, Martin:2017zxs}, for which our result could be generalised.
\subsection{Correlator}
Plugging \Eq{eq:Si:def} into \Eq{eq:correlator:def}, one obtains for the two-time correlator
\bea
	E(t_a,t_b)&\equiv&\frac{1}{2}\Braket{0,0| \left\{\hat S_1(t_a),\hat S_2(t_b)\right\}  |0,0}\\
	&=&\Re\mathrm e \Braket{0,0| \hat U^\dagger(t_a)\hat S_1\hat U(t_a)\hat U^\dagger(t_b)\hat S_2\hat U(t_b)  |0,0}\\
	&=&\sum_{n=-\infty}^\infty\sum_{m=-\infty}^\infty (-1)^{n+m}
	\int_{n\ell}^{(n+1)\ell}\dd \tilde Q_1 \int_{m\ell}^{(m+1)\ell}\dd \bar Q_2\nonumber \\ & &
	\Rea \left[ \Braket{0,0| \hat U^\dagger(t_a)\Ket{\tilde{Q}_1}\Bra{\tilde{Q}_1}\hat U(t_a)\hat U^\dagger(t_b) \Ket{\bar{Q}_2}\Bra{\bar{Q}_2}\hat U(t_b)  |0,0}\right]\, .
\eea
By introducing the closure relation $\hat{1}=\int_{-\infty}^\infty \dd \tilde{Q}_2 \Ket{\tilde{Q}_2}\Bra{\tilde{Q}_2}$ between the first and the second correlators in the argument of the real part, and $\hat{1}=\int_{-\infty}^\infty \dd \bar{Q}_1 \Ket{\bar{Q}_1}\Bra{\bar{Q}_1}$ between the second and the third correlators, one obtains
\bea
E(t_a,t_b)&=&\sum_{n=-\infty}^\infty\sum_{m=-\infty}^\infty (-1)^{n+m}
	\int_{n\ell}^{(n+1)\ell}\dd \tilde Q_1 \int_{m\ell}^{(m+1)\ell}\dd \bar Q_2
	\int_{-\infty}^{\infty}\dd \tilde Q_2 \int_{-\infty}^{\infty}\dd \bar Q_1 \nonumber \\
	&& \Re\mathrm e\com{\Psi_\mathrm{2sq}^*(\tilde Q_1,\tilde Q_2;t_a)\Psi_\mathrm{2sq}(\bar Q_1,\bar Q_2;t_b)
	\Braket{\tilde Q_1, \tilde Q_2|\hat U(t_a) \hat U^\dagger(t_b)| \bar Q_1, \bar Q_2} }.
	\label{Eab_detail}
\eea
In this expression, the wavefunction in position space is given by \Eq{2sq_wave_func}, as derived from \Eq{2sq} in \App{App:ket:tmss}. It has a Gaussian structure. The correlator $\Braket{\tilde Q_1, \tilde Q_2|\hat U(t_a) \hat U^\dagger(t_b)| \bar Q_1, \bar Q_2}$ is calculated in \App{app:correlation}, where it is shown to be also of the Gaussian form, see \Eq{QQUUQQ}. Combining these two results, one obtains
\bea
\hspace{-5mm}
	\Psi_\mathrm{2sq}^*(\tilde Q_1,\tilde Q_2;t_a)\Psi_\mathrm{2sq}(\bar Q_1,\bar Q_2;t_b)
	\Braket{\tilde Q_1, \tilde Q_2|\hat U(t_a) \hat U^\dagger(t_b)| \bar Q_1, \bar Q_2}
	=\mathcal A\exp\left( \frac{1}{2} X^\mathrm{T}\Lambda X  \right),
	\label{lambda_expression}
\eea
where $X^{\mathrm{T}} = \com{\bar Q_1,\bar Q_2,\tilde Q_1,\tilde Q_2}$, 
\bea
	\mathcal A = \left(\pi^2\cosh^2r_a\cosh^2r_b
	\sqrt{1-e^{4i\varphi_a}\tanh^2r_a}\,\sqrt{1-e^{-4i\varphi_b}\tanh^2r_b} \,\sqrt{\det M}
	\right)^{-1},
	\label{calA}
\eea
with $\det M$ given in \Eq{detM}, and
\bea
	\Lambda = \com{
	\begin{array}{cccc}
		\mathcal D_1 & \mathcal D_2 & D_3 & D_4 
		\vspace{1mm}\\
		 \mathcal D_2 & \mathcal D_1 & D_4 & D_3 
		\vspace{1mm}\\
		 D_3 & D_4 & \bar{\mathcal D}_1 & \bar{\mathcal D}_2 
		\vspace{1mm}\\
		 D_4 & D_3 & \bar{\mathcal D}_2   & \bar{\mathcal D}_1
	\end{array}
	} .  
	\label{lambda_def}
\eea
In this last expression, we have introduced 
\bea
	\mathcal D_1&=&\frac{1}{2} + A(r_b,\varphi_b) - D_1, \quad
	\bar{\mathcal D}_1=\frac{1}{2} + A^*(r_a,\varphi_a) - \bar{D}_1\\
	\mathcal D_2&=&B(r_b,\varphi_b) - D_2 , \quad
	\bar{\mathcal D}_2=B^*(r_a,\varphi_a) - \bar D_2
	,
\eea
where $A$ and $B$ are given in \Eq{AB}, and $D_1$, $D_2$, $\bar{D}_1$ and $\bar{D}_2$ by \Eqs{eq:Di:def} and~\eqref{eq:Dibar:def}. 

The Gaussian integral over $\bar Q_1$ and $\tilde Q_2$ in \Eq{Eab_detail} can then be performed, and one obtains
\bea
	E(t_a,t_b)&=&\Re\mathrm e \com{ \frac{2\pi \mathcal A}{\sqrt{\mathcal D_1\bar{\mathcal D}_1 - D_4^2} }
	\sum_{n,m=-\infty}^\infty
	(-1)^{n+m} \int_{n\ell}^{(n+1)\ell}\dd Y_1 \int_{m\ell}^{(m+1)\ell}\dd Y_2
	\exp\left(\frac{1}{2} Y^\mathrm{T}\Xi Y  \right)  },
	\nonumber \\
	\label{analytic_rlt}
\eea
where
\bea
	\Xi_{11} &=& \mathcal D_1 - \frac{\bar{\mathcal D}_1\mathcal D_2^2 + \mathcal D_1D_3^2 - 2\mathcal D_2D_3D_4   }{\mathcal D_1\bar{\mathcal D}_1 - D_4^2}   \label{def_Xi11}, \\
	\Xi_{22} &=& \bar{\mathcal D}_1 - \frac{\mathcal D_1\bar{\mathcal D}_2^2 + \bar{\mathcal D}_1D_3^2 - 2\bar{\mathcal D}_2D_3D_4   }{\mathcal D_1\bar{\mathcal D}_1 - D_4^2} ,  \label{def_Xi22} \\
	\Xi_{12} &=&\Xi_{21}=  D_4 - \frac{\bar{\mathcal D}_1\mathcal D_2D_3 + \mathcal D_1\bar{\mathcal D}_2D_3 - \mathcal D_2\bar{\mathcal D}_2D_4 - D_3^2D_4   }{\mathcal D_1\bar{\mathcal D}_1 - D_4^2}.   \label{def_Xi12}
\eea

A few comments are in order regarding this formula. First, in \App{app:Gaussian_int}, we study the conditions under which the Gaussian integrals in \Eq{analytic_rlt} are convergent, and we find that it is the case if
\bea
	\Rea \prn{\Xi_{11}}<0 ,\quad
	\Rea \prn{\Xi_{22}}<0 ,\quad
	\Rea \prn{\Xi_{11}-\frac{\Xi_{12}^2}{\Xi_{22}}}<0,\quad
	\Rea \prn{\Xi_{22}-\frac{\Xi_{12}^2}{\Xi_{11}}}<0,
	\label{integrability_refined}
\eea
see \Eq{Gauss_int_cond_refined}. In practice, we have checked that these conditions are always satisfied in all physical configurations we have studied, but let us stress that all formulas derived below assume that \Eq{integrability_refined} holds. Second, as explained in footnote~\ref{foot_sqrt_detM} in \App{app:correlation}, in order to solve the sign ambiguity in the term $\sqrt{\det M}$ that appears in \Eq{calA} for $\mathcal A$, one needs to solve an eigenvalue problem for $M$, which is a $12\times 12$ matrix. However, one can show that the following identity holds for any $r_a$, $r_b$, $\varphi_a$, $\varphi_b$, $\theta_a$ and $\theta_b$,
\bea
	\frac{\mathcal A}{\sqrt{\mathcal D_1\bar{\mathcal D}_1 - D_4^2}} = \frac{\sqrt{\Xi_{11}\Xi_{22}- \Xi_{12}^2 } }{4\pi^2}\, ,
	\label{calA_const}
\eea
which simplifies the prefactor in \Eq{analytic_rlt} and allows us to avoid dealing with the eigenvalue problem.\footnote{Contrary to $\sqrt{\det M}$, $\sqrt{\det \Xi}$ does not require to solve an eigenvalue problem. Indeed, since $\Xi$ is a $2\times 2$ matrix, it has two eigenvalues with a negative real sign (see below), which can be written as $d_\pm = - \rho_\pm \ee^{i\alpha_{\pm}}$, with $\rho_{\pm}>0$ and $-\pi/2<\alpha_\pm<\pi/2$. This means that $\det\Xi = d_+ d_- = \rho_+ \rho_- \ee^{i (\alpha_+ + \alpha_-)}$. Since $-\pi<\alpha_++\alpha_-<\pi$, the phase of $\det\Xi$ never crosses the branch cut of the square root function and thus it leaves no sign ambiguity. 
Notice that, strictly speaking, in order to ensure that the eigenvalues of $\Xi$ have a negative real part, one needs to impose a stronger condition than \Eq{integrability_refined}, where the last two inequalities are replaced with $\Rea\prn{\Xi_{11}} \Rea\prn{\Xi_{22}}-\Rea\prn{\Xi_{12}}^2>0$. This is the analogue of \Eq{abs_convergence_cond}, while \Eq{integrability_refined} is the analogue of \Eq{Gauss_int_cond_refined}. 
Furthermore, in order to ensure that $\Xi$ can be diagonalised 
by an orthogonal matrix, one has to verify that $\Xi$ is normal \ie $\Xi\Xi^\dagger = \Xi^\dagger\Xi$, see also footnote~\ref{foot_sqrt_detM}.
From \Eq{LargeSqueezingXi}, one can check that this is the case when $r_a = r_b$ in the large-squeezing limit.
Otherwise, one can nevertheless find an invertible matrix that transforms $\Xi$ into a diagonal matrix in the following way.
If one writes $\Xi = -A + iB$, as long as $ = - \Rea(\Xi)$ is positive definite, $\sqrt A$ is well-defined. 
Since $\sqrt A^{-1} B \sqrt A^{-1}$ is real and symmetric, it can be diagonalised by an orthogonal matrix $O$.
Then, one can construct an invertible matrix $V\equiv \sqrt A^{-1} O$, so that $V^\mathrm{T} \Xi V$ is diagonal.
Although its diagonal elements are not the eigenvalues of $\Xi$ anymore, this ``diagonalisation" makes it possible to perform the Gaussian integral analytically.
In any case, the procedure is straightforward since one only needs to compute the eigenvalues of a $2\times 2$ matrix.
}
Finally, let us point out that although $E(t_a,t_b)$ a priori depends on 6 parameters, $r_a$, $r_b$, $\varphi_a$, $\varphi_b$, $\theta_a$ and $\theta_b$, only the difference between the rotation angles is involved, namely $\Delta\theta\equiv\theta_a-\theta_b$, see the formulas obtained in \App{app:correlation}. This is because, as mentioned above, the rotation angle only appears in the time evolution from $t_a$ and $t_b$. As a consequence, $E(t_a,t_b)$ depends on 5 parameters only.
\section{Analytical limits}
\label{sec:Analytical:Limits}
Before evaluating \Eq{analytic_rlt} numerically, and exploring whether or not there are configurations where the Bell inequality~\eqref{eq:Bell:inequality} can be violated, let us first study some limiting cases analytically. This will be useful to design a strategy for the numerical exploration of \Sec{sec:Numerical:Exploration}, which is otherwise tedious given the large dimensionality of parameter space. Indeed, at each time $t_a$, $t_a'$, $t_b$, $t_b'$, one must specify a squeezing amplitude, a squeezing angle and a rotation angle. Because of the above remark on the rotation angles, only the combinations $\Delta\theta_{ab}$, $\Delta\theta_{ab'}$, $\Delta\theta_{a'b}$ and $\Delta\theta_{a'b'}$ are involved, which are related through the Chasles relation $\Delta\theta_{ab}-\Delta\theta_{a'b}+\Delta\theta_{a'b'}-\Delta\theta_{ab'}=0$. This still leaves us with 11 parameters to explore, and at each point in parameter space, each correlator appearing in \Eq{eq:Bell:inequality} is given by a double infinite sum of double integrals. Even though one of the two integrals can be performed analytically, see \Eq{E12_2D_GaussInt2} in \Sec{sec:Numerical:Exploration} below, this remains computationally heavy, which justifies the need for further analytical insight.  
\subsection{Large-$\ell$ limit}
\label{sec:large:l:limit}
In the limit where $\ell$ is large, as mentioned above the spin operator~\eqref{eq:Si:def} becomes the sign operator, see \Eq{eq:Spin:Large:L:sign}. In this regime, only four terms in the sum~\eqref{analytic_rlt} remain, namely those for $(n,m)=(0,0)$, $(-1,0)$, $(0,-1)$ and $(-1,-1)$. By performing the change of integration variable $Y_1\to- Y_1$ and $Y_2\to -Y_2$, one can see that the integrals for $(n,m)=(0,0)$ and $(n,m)=(-1,-1)$ are the same, and that the integrals for $(n,m)=(-1,0)$ and $(n,m)=(0,-1)$ are the same. This gives rise to
\bea
	E(t_a,t_b)
	&=&\Re\mathrm e \left[ \frac{4\pi \mathcal A}{\sqrt{\mathcal D_1\bar{\mathcal D}_1 - D_4^2} }
	 \prn{  \int_{0}^{\infty}\dd Y_1 \int_{0}^{\infty}\dd Y_2 
	 - \int_{-\infty}^{0}\dd Y_1 \int_{0}^{\infty}\dd Y_2 }
	\exp\left( \frac{1}{2} Y^\mathrm{T}\Xi Y  \right)   \right]
	.
	\nonumber \\
	\label{eq:LargeL:1}
\eea
In \App{app:Gaussian_int}, it is shown how the two integrals appearing in \Eq{eq:LargeL:1} can be expressed in terms of the $\mathrm{arc}\tan$ function, see \Eqs{Gauss_int_1} and~\eqref{Gauss_int_2}. Making use of \Eq{calA_const}, this leads to
\bea
	E(t_a,t_b)	 \underset{\ell\to\infty}{\longrightarrow} \frac{2}{\pi }\Rea  \left[
	\mathrm{arc}\tan\prn{ \frac{\Xi_{12}}{\sqrt{\Xi_{11}\Xi_{22}- \Xi_{12}^2 }}  }
	   \right] .
	\label{Eab_Bell_analytic}
\eea
This formula~\eqref{Eab_Bell_analytic} is compared with a full numerical computation of \Eq{analytic_rlt} in \Fig{fig:small_ell}, where one can check that it correctly reproduces the asymptotic value of $E(t_a,t_b)	$ when $\ell\to\infty$.
\subsection{Small-$\ell$ limit}
\label{sec:small:ell:limit}
\begin{figure}[t]
\begin{center}
\includegraphics[width=0.496\textwidth]{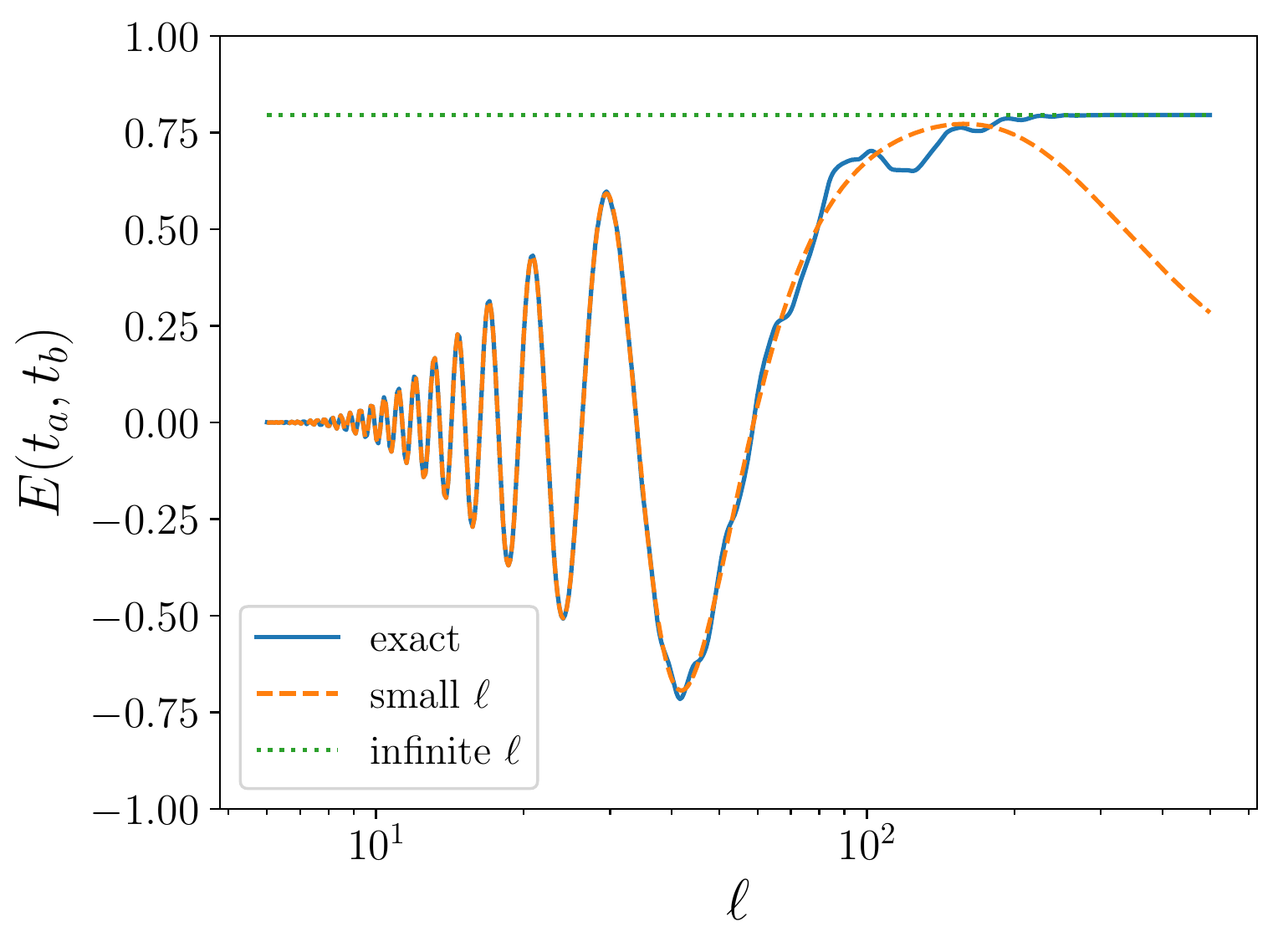}
\caption{The correlation function $E(t_a, t_b)$ for $r_a = r_b = 5, \varphi_a = -0.2, \varphi_b = 0.2, \Delta\theta = 0.5$, as a function of $\ell$. The blue line corresponds to the numerical computation of \Eq{analytic_rlt}, the dashed orange line to the small-$\ell$ approximation~\eqref{E12_small_ell}, and the dotted green line to the large-$\ell$ approximation~\eqref{Eab_Bell_analytic}. One can check that these two approximations give good fits to the full result in their respective domains of validity, $\ell\ll \ee^r$ and $\ell\gg \ee^r$.}  
\label{fig:small_ell}
\end{center}
\end{figure}
In the small-$\ell$ limit, conversely, an infinitely large number of terms in the sum over $n$ and $m$ in \Eq{analytic_rlt} substantially contribute to the result. However, the range of the integrals appearing in \Eq{analytic_rlt} becomes very small in that limit, so one can Taylor expand the integrand in each range and perform the integral analytically. This procedure is nonetheless delicate and in \App{app:small_ell}, it is performed in detail, making use of elliptic theta functions to carefully resum and expand the different contributions. Plugging \Eqs{eq:app:low:l:result} and~\eqref{calA_const} into \Eq{analytic_rlt}, one obtains
\bea
E(t_a,t_b) \underset{\ell \ll e^r}{\longrightarrow}
\frac{8}{\pi^2}
 \Rea \left(e^{p_+}-e^{p_-}\right)\, ,
 \quad \quad\quad\mathrm{where} \quad\quad
 p_\pm\equiv \frac{\pi^2\prn{\Xi_{11}+\Xi_{22} \pm 2\Xi_{12}}}{2\prn{\Xi_{11}\Xi_{22} - \Xi_{12}^2}\ell^2}\, .
 \label{E12_small_ell}
\eea
Notice that the expansion performed in \App{app:small_ell} requires that $\Rea(\Xi_{11} - \Xi_{12}^2/\Xi_{22})<0$, see \Eq{eq:expansion:theta4:alpha}, which here is guaranteed by \Eq{integrability_refined}. Note also that ``$\ell \ll \ee^r$'' is a shorthand notation for $\ell \ll \umin(\ee^{r_a}, \ee^{r_b})$.
The formula~\eqref{E12_small_ell} is again compared with the full numerical computation of \Eq{analytic_rlt} in \Fig{fig:small_ell}, where one can check that it gives a very good fit to the full result, up to values of $\ell$ of order $\ee^r$. 
\subsection{Large-squeezing limit}
\label{sec:large:squeezing:limit}
Another limit of interest is when the squeezing amplitude of the state is large. 
A large amount of squeezing is associated to a large entanglement entropy and a large quantum discord between the two subsystems~\cite{Martin:2015qta}, \ie to the presence of genuine quantum correlations. In \Refa{Martin:2016tbd}, it was shown that the usual Bell inequalities can be violated by two-mode squeezed states  provided the squeezing amplitude is large enough (namely $r\gtrsim 1.2$), so one might expect that bipartite temporal Bell inequalities also require a minimum amount of squeezing.

Note that since the squeezing amplitude $r$ always appears in the form of $e^r$, the large-squeezing regime corresponds to $e^r \gg 1$ (hence the value $r = 5$, which is used in most numerical applications below, falls in that regime).
For convenience, we thus introduce the notation $u\equiv e^{-r}$, so the large-squeezing limit stands for $u_a,\, u_b\ll 1$. 
In this regime, from \Eqs{def_Xi11}-\eqref{def_Xi12}, one obtains
\bea
	\Xi_{11}&\simeq& -2 \frac{u_b^2 }{ \mathcal X}\, , \quad\quad  \label{Xi11_Squeeze} 
	\Xi_{22}\simeq -2 \frac{u_a^2 }{ \mathcal X} \, , \quad\quad 
	\Xi_{12}\simeq e^{i\Delta\theta} \prn{e^{2i\varphi_a} + e^{-2i\varphi_b}} \frac{u_a u_b}{ \mathcal X} \, , \quad \quad 
	\label{LargeSqueezingXi}
\eea
where
\bea
	\mathcal X
	= \frac{1}{8}\com{4 - e^{2i\Delta\theta} \prn{e^{2i\varphi_a} + e^{-2i\varphi_b}}^2}
	.  \label{ChiTilde_Squeeze} 
\eea
One can easily check that $\Rea \prn{\mathcal X} \geq 0$, so $\Rea(\Xi_{11})<0$ and $\Rea(\Xi_{22})<0$, and the two first conditions of  \Eq{integrability_refined} are satisfied. 
Moreover, since $\Xi_{11}-\Xi_{12}^2 / \Xi_{22} = -4u_b^2$, $\Xi_{22}-\Xi_{12}^2 / \Xi_{11} = -4u_a^2$, they are real and negative, which shows that the two last conditions of \Eq{integrability_refined} are satisfied too.
\subsection{Large-$\ell$, large-squeezing limit}
\label{sec:large:ell:large:squeezing:limit}
\begin{figure}[t]
\begin{center}
\includegraphics[width=0.496\textwidth]{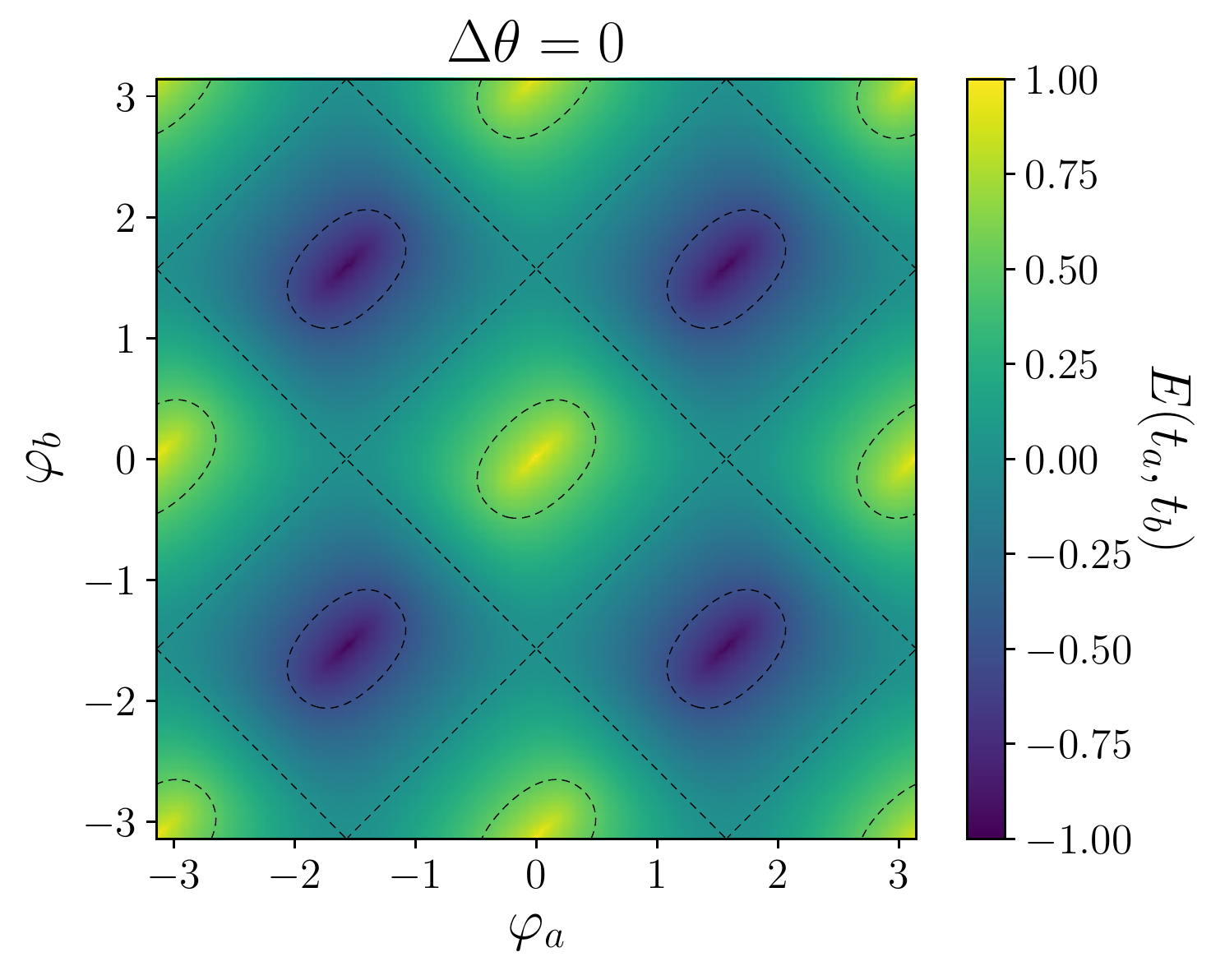}
\includegraphics[width=0.496\textwidth]{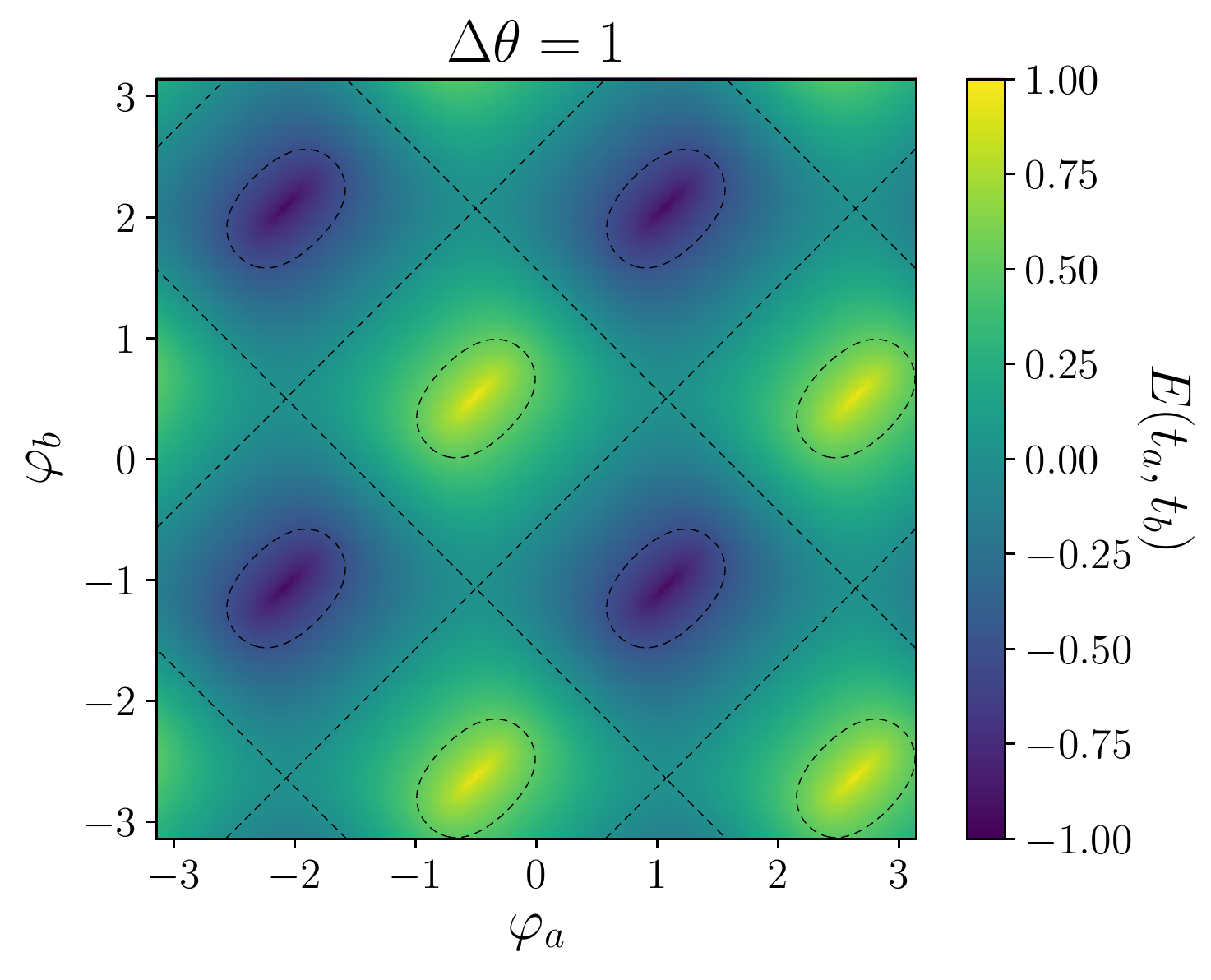}
\caption{The correlation function $E(t_a, t_b)$ in the large-$\ell$, large-squeezing limits, \ie $\ell,\, r_a,\, r_b\rightarrow\infty$, as a function of $\varphi_a$ and $\varphi_b$, computed with \Eq{large_r_NoInf_refined}. The black dashed lines are contour lines for $E(t_a,t_b) = 0.5, 0$ and $-0.5$ and are guide for the eye. In the left panel, $\Delta\theta = 0$, while $\Delta\theta = 1$ in the right panel. These two figures are simply related by a translation: when going from the left to the right panel, $\varphi_a$ is shifted by $-\Delta\theta/2$ and  $\varphi_b$  by $\Delta\theta/2$.}  
\label{fig:E12_Squeeze_map}
\end{center}
\end{figure}
Let us now combine the two limits studied in \Secs{sec:large:l:limit} and~\ref{sec:large:squeezing:limit}, and investigate the large-$\ell$, large-squeezing limit. By plugging \Eqs{Xi11_Squeeze} and~\eqref{ChiTilde_Squeeze} into \Eq{Eab_Bell_analytic}, one obtains
\begin{eqnarray}
\label{large_r_NoInf_refined}
	E(t_a,t_b)\underset{\ell,r_a,r_b\to\infty}{\xrightarrow{\hspace*{12mm}}} 
	\frac{2}{\pi }\Rea \left\lbrace
	\mathrm{arc}\tan\left[ \frac{e^{i\Delta\theta} \prn{e^{2i\varphi_a} + e^{-2i\varphi_b}}}{\sqrt{4 - e^{2i\Delta\theta} \prn{e^{2i\varphi_a} + e^{-2i\varphi_b}}^2}}  \right]\right\rbrace   \, .
\end{eqnarray}
This formula is displayed in \Fig{fig:E12_Squeeze_map}, as a function of $\varphi_a$ and $\varphi_b$, for $\Delta\theta=0$ (left panel) and $\Delta\theta=1$ (right panel). The black dashed lines are contour lines for $E(t_a,t_b) = 0.5, 0$ and $-0.5$, and are guide for the eye. As can be seen from \Eq{large_r_NoInf_refined}, when $\Delta\theta$ varies the map is simply modified by a translation, where $\varphi_a$ is shifted by $-\Delta\theta/2$ and  $\varphi_b$  by $\Delta\theta/2$.

The correlation is maximal, \ie $E(t_a,t_b)=\pm 1$, when the denominator of the argument of the $\mathrm{arc}\tan$ function in \Eq{large_r_NoInf_refined} vanishes. This happens when $\varphi_a=(n \pi-\Delta\theta)/2$ and $\varphi_b=(m \pi+\Delta\theta)/2$, where $n$ and $m$ are two integers of the same parity. More precisely, $E(t_a,t_b)=1$ when $n$ and $m$ are even, and $E(t_a,t_b)=-1$ when $n$ and $m$ are odd.

\begin{figure}[t]
\begin{center}
\includegraphics[width=0.5\textwidth]{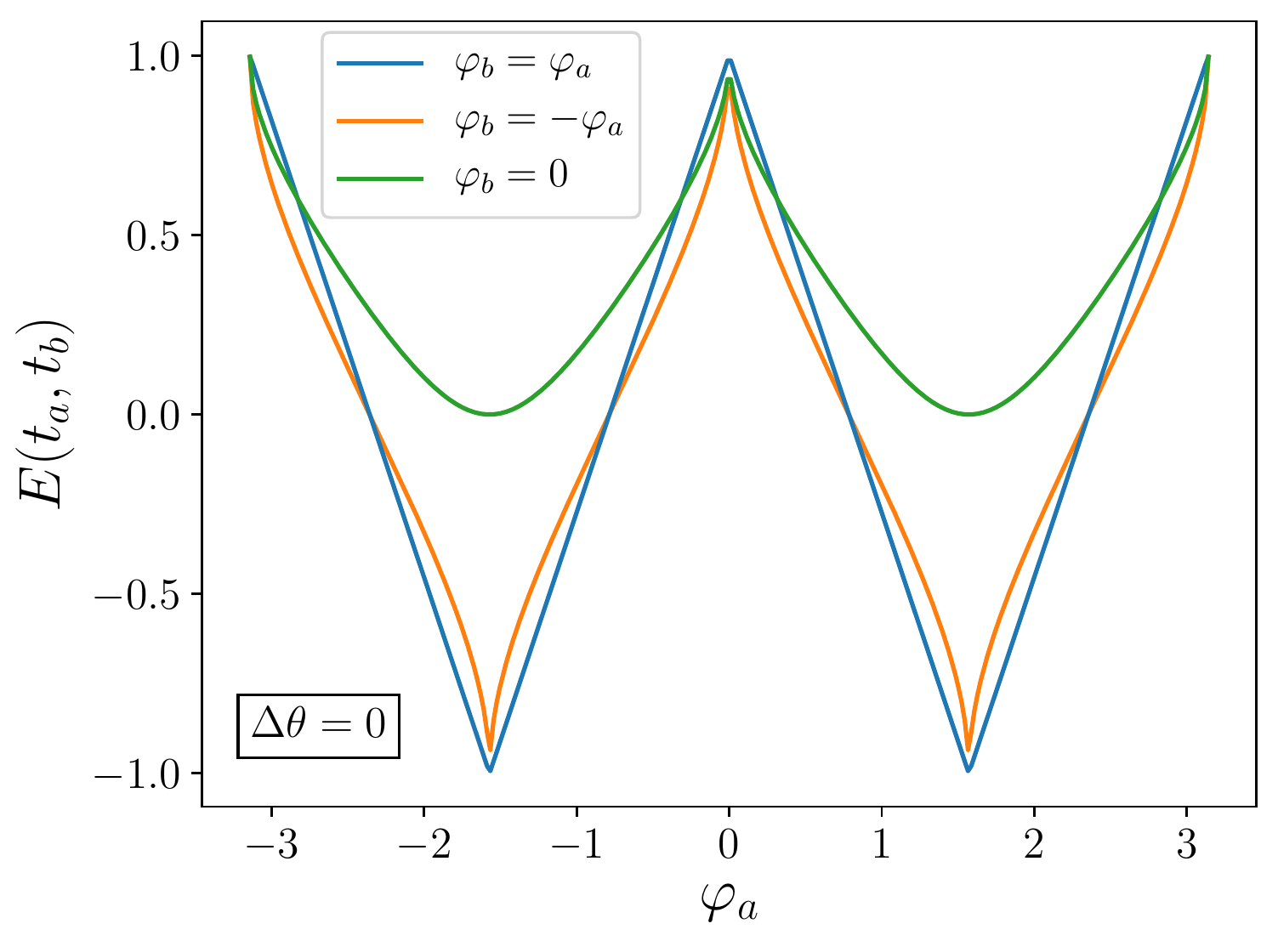}
\caption{The correlation function $E(t_a, t_b)$ in the large-$\ell$, large-squeezing limits, \ie $\ell,\, r_a,\, r_b\rightarrow\infty$, as computed in \Eq{large_r_NoInf_refined}, as a function of $\varphi_a$ for  $\Delta\theta = 0$ and $\varphi_b=\varphi_a$ (blue line), $\varphi_b=-\varphi_a$ (orange line), and $\varphi_b=0$ (green line). These are slices from the left panel of \Fig{fig:E12_Squeeze_map}. For $\varphi_b=\varphi_a$, $E(t_a, t_b)$ is a piecewise affine function of $\varphi_a$, see \Eq{eq:E:linear}. For $\varphi_b=-\varphi_a$, the derivative of $E(t_a, t_b)$ diverges at the points where $E(t_a,t_b)=\pm1$, see \Eq{eq:E:cusp:point}.}  
\label{fig:E12_Squeeze_curve}
\end{center}
\end{figure}
In \Fig{fig:E12_Squeeze_curve} we show three slices from the maps displayed in \Fig{fig:E12_Squeeze_map}, where $E(t_a,t_b)$ is plotted as a function of $\varphi_a$ for different choices of $\varphi_b$ and $\Delta\theta=0$. The blue line corresponds to $\varphi_b=\varphi_a$, and one can see that $E(t_a,t_b)$ is a piecewise affine function of $\varphi_a$. Indeed, more generally, the lines $\varphi_b = \varphi_a + \Delta\theta + n\pi \, (n\in  \mathbb Z)$ connect all the points where $E(t_a,t_b)=\pm 1$, and along these lines one has
\bea
\label{eq:E:linear}
	E(t_a,t_b) = (-1)^m\times
	\begin{cases}
		-\frac{2}{\pi}\prn{\varphi_a + \varphi_b - 2n\pi} + 1 & \mathrm{for} \quad
		2n\pi < \varphi_a + \varphi_b \leq (2n+1) \pi \\
		\frac{2}{\pi}\prn{\varphi_a + \varphi_b - 2n\pi} + 1 & \mathrm{for} \quad
		(2n-1)\pi < \varphi_a + \varphi_b \leq 2n \pi
	\end{cases}\, .
\eea
The orange line corresponds to $\varphi_b=-\varphi_a$, where one notices the existence of cusps at the points where $E(t_a,t_b)=\pm1$.
Around the cusps, $E(t_a,t_b)$ indeed behaves as
\begin{eqnarray}
\label{eq:E:cusp:point}
	E(t_a,t_b)\simeq \pm1 \pm \frac{2\sqrt{2\prn{\varphi_a-\varphi_{a,\mathrm{cusp}}}}}{\pi} \, ,
\end{eqnarray}
where $\varphi_{a,\mathrm{cusp}}$ indicates the location of a cusp point. Finally, the green line stands for a fixed value of $\varphi_b$, namely $\varphi_b=0$.
\section{Numerical exploration}
\label{sec:Numerical:Exploration}
Beyond the limits studied in the previous section, one has to compute \Eq{analytic_rlt} numerically. It is useful to notice that one of the two integrals appearing in \Eq{analytic_rlt} can be performed in terms of the complementary error function $\mathrm{erfc}(z)\equiv 1-\erf(z)$, where the error function $\erf(z)$ is defined below \Eq{eq:app:Gaussia:int:j:def}. For $\Xi_{22}\neq0$, one has
\begin{eqnarray}
	E(t_a,t_b)
	 &=&\Rea  \Bigg\{ - \frac{\sqrt{\Xi_{11}\Xi_{22}- \Xi_{12}^2 } }{2\sqrt{2\pi}\sqrt{-\Xi_{22}} }
	\sum_{n=-\infty}^\infty\sum_{m=-\infty}^\infty (-1)^{n+m}
	\int_{n\ell}^{(n+1)\ell}\dd Y_1  \nonumber \\
	&& \quad \left( \mathrm{erfc}\left\lbrace \sqrt{-\frac{\Xi_{22}}{2}} \left[(m+1)\ell + \frac{\Xi_{12}}{\Xi_{22}} Y_1\right]\right\rbrace  - \mathrm{erfc}\left[\sqrt{-\frac{\Xi_{22}}{2}} \prn{m\ell + \frac{\Xi_{12}}{\Xi_{22}} Y_1}\right]  \right) \nonumber \\
	&&\quad \exp\com{\frac{1}{2}\prn{\Xi_{11} - \frac{\Xi_{12}^2}{\Xi_{22}}} Y_1^2}
	\Bigg\} 
	\label{E12_2D_GaussInt2}
\end{eqnarray}
where \Eq{calA_const} has been used to simplify the prefactor.

This expression can be further simplified (for the sake of numerical computation) by noticing that \Eq{analytic_rlt} is of the form $\sum_{n = -\infty}^\infty\sum_{m = -\infty}^\infty a_{n,m}$ where, by performing the change of integration variable $Y_1\to -Y_1$ and $Y_2\to -Y_2$, one has $a_{-n-1,-m-1} = a_{n,m}$. One can use this relation to restrict the sum over positive values of $m$ only, and replace $\sum_{m = -\infty}^\infty$ with $2\sum_{m = 0}^\infty$. The sum over $m$ can then be re-ordered since the complementary error functions are absolutely convergent as long as $\Re\mathrm e(\Xi_{22})<0$, which is required according to \Eq{integrability_refined}. This gives rise to
\bea
 \label{E12_finite_ell_numerical}
	 E(t_a,t_b)&=&\Re\mathrm e \Bigg\{
	 \frac{\sqrt{\Xi_{11}\Xi_{22}- \Xi_{12}^2 } }{\sqrt{2\pi} \sqrt{-\Xi_{22}}}
	  \sum_{n=-\infty}^\infty (-1)^n \int_{n\ell}^{(n+1)\ell}\dd Y 
	 \nonumber \\
	&& \quad  \left\lbrace  2 \sum_{m=0}^\infty (-1)^{m}
	   \mathrm{erfc}\left[\sqrt{\frac{-\Xi_{22}}{2}} \prn{m\ell + \frac{\Xi_{12}}{\Xi_{22}} Y}\right]
	 - \mathrm{erfc}\prn{\frac{-\Xi_{12}}{\sqrt{-2\Xi_{22}}} Y}\right\rbrace \nonumber \\
	&& \quad \exp\com{\frac{1}{2}\prn{\Xi_{11} - \frac{\Xi_{12}^2}{\Xi_{22}}} Y^2}
	 \Bigg\} . 
\eea
This expression is helpful to compute $E(t_a,t_b)$ numerically, and in practice, we truncate the sums over $n$ and $m$ at an order above which we check that the dependence of the result on the truncation order becomes negligible.
\begin{figure}[t]
\begin{center}
\includegraphics[width=0.52\textwidth]{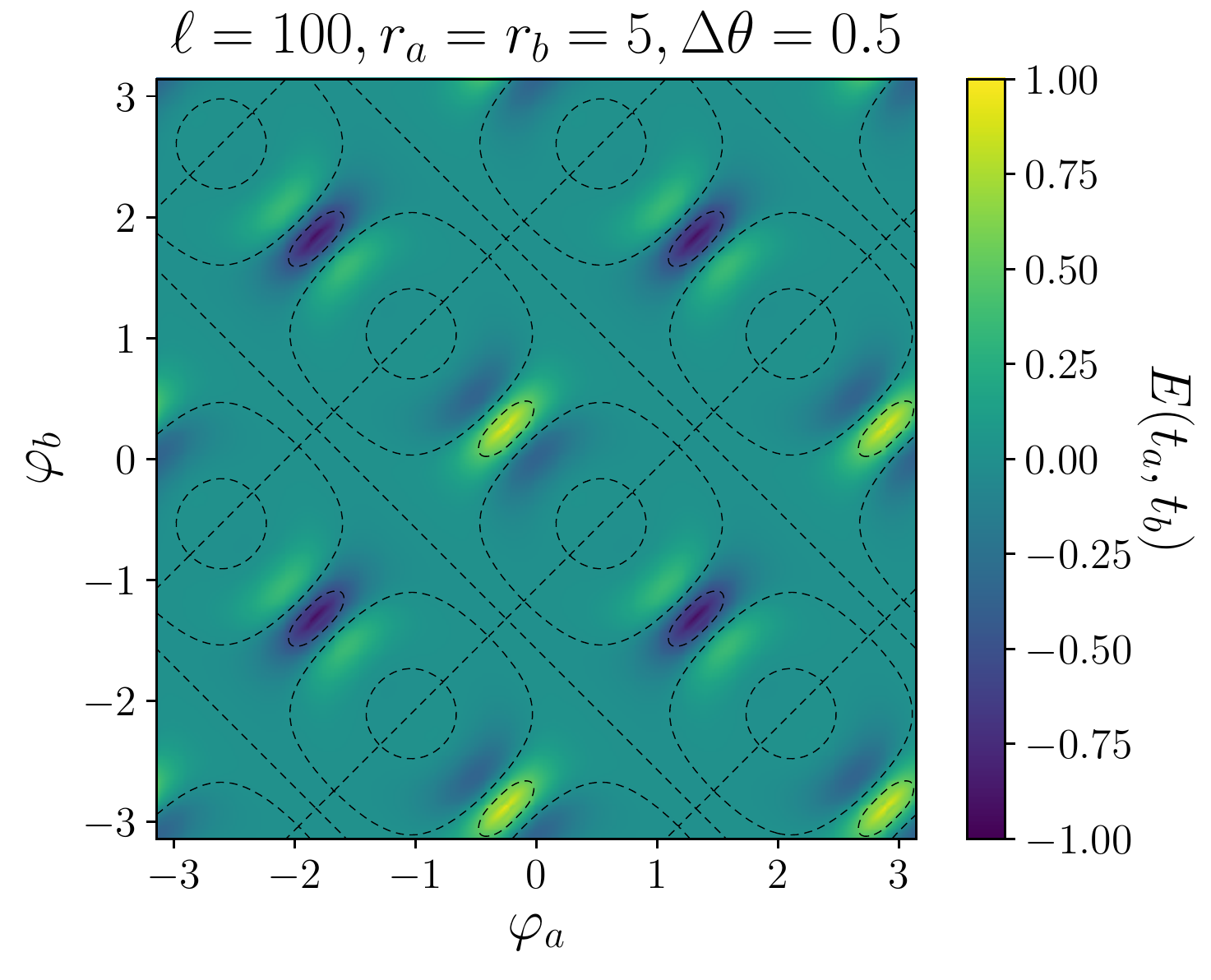}
\includegraphics[width=0.47\textwidth]{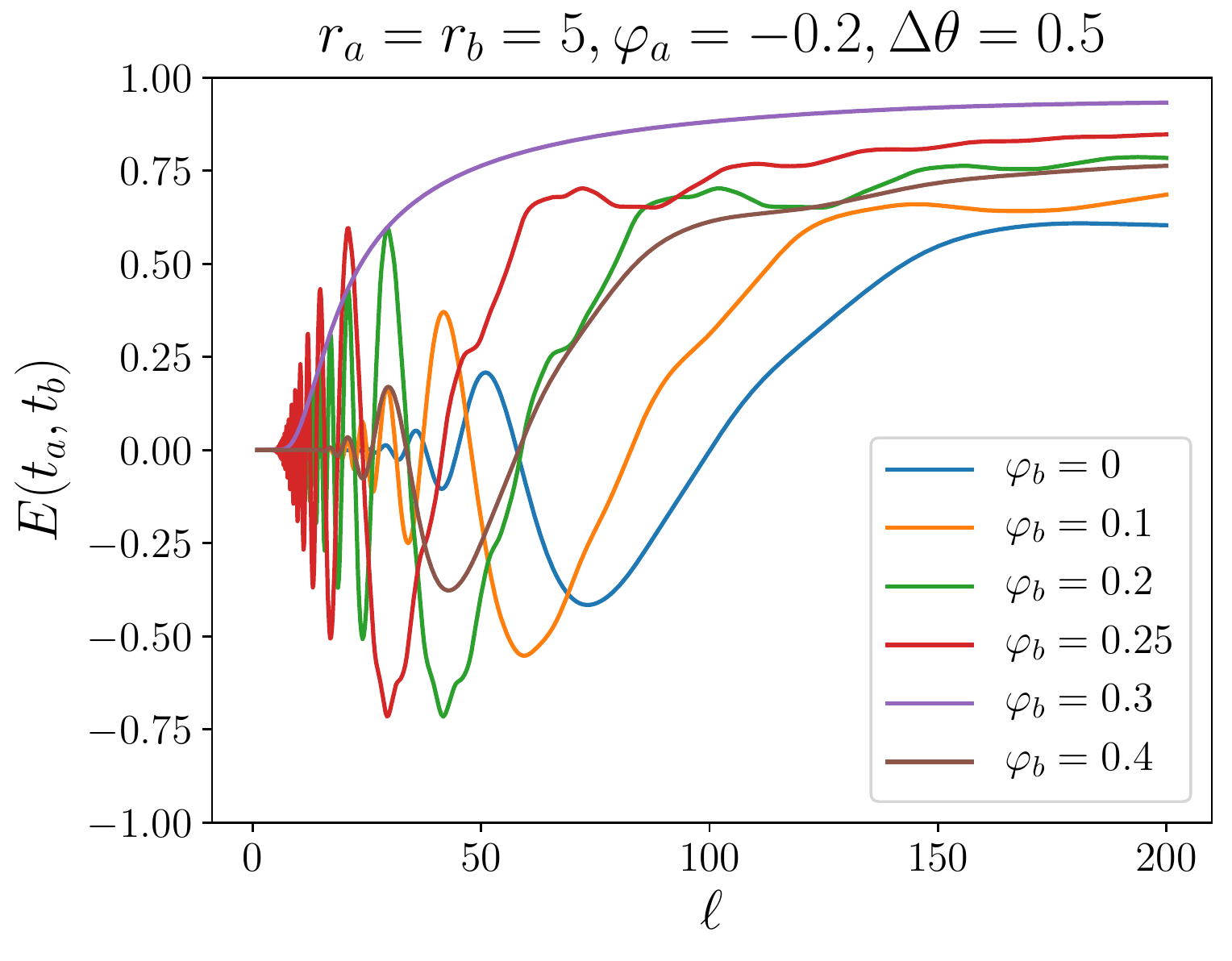}
\caption{Left panel: $E(t_a,t_b)$ for $\ell = 100$, $r_a = r_b = 5$ and $\Delta\theta = 0.5$, as a function of $\varphi_a$ and $\varphi_b$.  The black dashed lines are contours for $E(t_a,t_b) = 0.5$, $0$ and $-0.5$.
	Right panel: $E(t_a,t_b)$ as a function of $\ell$ for $r_a = r_b = 5$, $\varphi_a = -0.2$, $\Delta\theta = 0.5$. Different colours label different values of $\varphi_b$.}  
\label{fig:E12:Numerical}
\end{center}
\end{figure}

In the left panel of \Fig{fig:E12:Numerical}, the correlation function is displayed for $r_a=r_b=5$, $\Delta\theta=0.5$ and $\ell=100$, as a function of $\varphi_a$ and $\varphi_b$.
From \Fig{fig:small_ell}, one can check that $\ell = 100$ is too large for the small-$\ell$ approximation developed in \Sec{sec:small:ell:limit} to apply, since it requires $\ell\ll \ee^{r}$, and too small for the large-$\ell$ approximation developed in \Sec{sec:large:l:limit} to apply, since it requires $\ell\gg \ee^{r}$. This value of $\ell$ is therefore ``intermediate'' in that sense. This is further confirmed by noting the difference between the left panel of \Fig{fig:E12:Numerical} and \Fig{fig:E12_Squeeze_map}, which displays the large-$\ell$ (and large squeezing, which here applies since $r=5$) limit. For intermediate $\ell$, one notices in the left panel of \Fig{fig:E12:Numerical} the presence of local maximums and local minimums, which do not exist in the limit where $\ell$ is infinite. As we will see below, those local extremums are crucial to obtain violations of bipartite temporal Bell inequalities.

In order to better depict the role played by the parameter $\ell$, which in principle is left to the free choice of the observer, in the right panel of \Fig{fig:E12:Numerical} we show the correlation function $E(t_a,t_b)$ as a function of $\ell$, for $r_a=r_b=5$, $\varphi_a=-0.2$ and $\Delta\theta=0.5$, for a few values of $\varphi_b$. This confirms the tendency observed in \Fig{fig:small_ell}: when $\ell\ll \ee^{r}$, \ie in the small-$\ell$ regime, there are oscillations, the amplitude and frequency of which strongly depend on the squeezing and rotation angles, while when $\ell\gg \ee^r$, one asymptotes the infinite-$\ell$ result. One should note that the case $\varphi_b = \varphi_a + \Delta\theta$ is an exception since no oscillation appears at small values of $\ell$ in that configuration (this corresponds to the violet curve, $\varphi_b=0.3$, in the right panel of \Fig{fig:E12:Numerical}). This is because, in the large-squeezing limit, all components of $\Xi$ are real, as can be seen from \Eqs{Xi11_Squeeze}-\eqref{ChiTilde_Squeeze}, while the oscillations come from evaluating exponential functions with complex arguments.

$ $\\
\par
\begin{figure}[t]
\begin{center}
\includegraphics[width=0.469\textwidth]{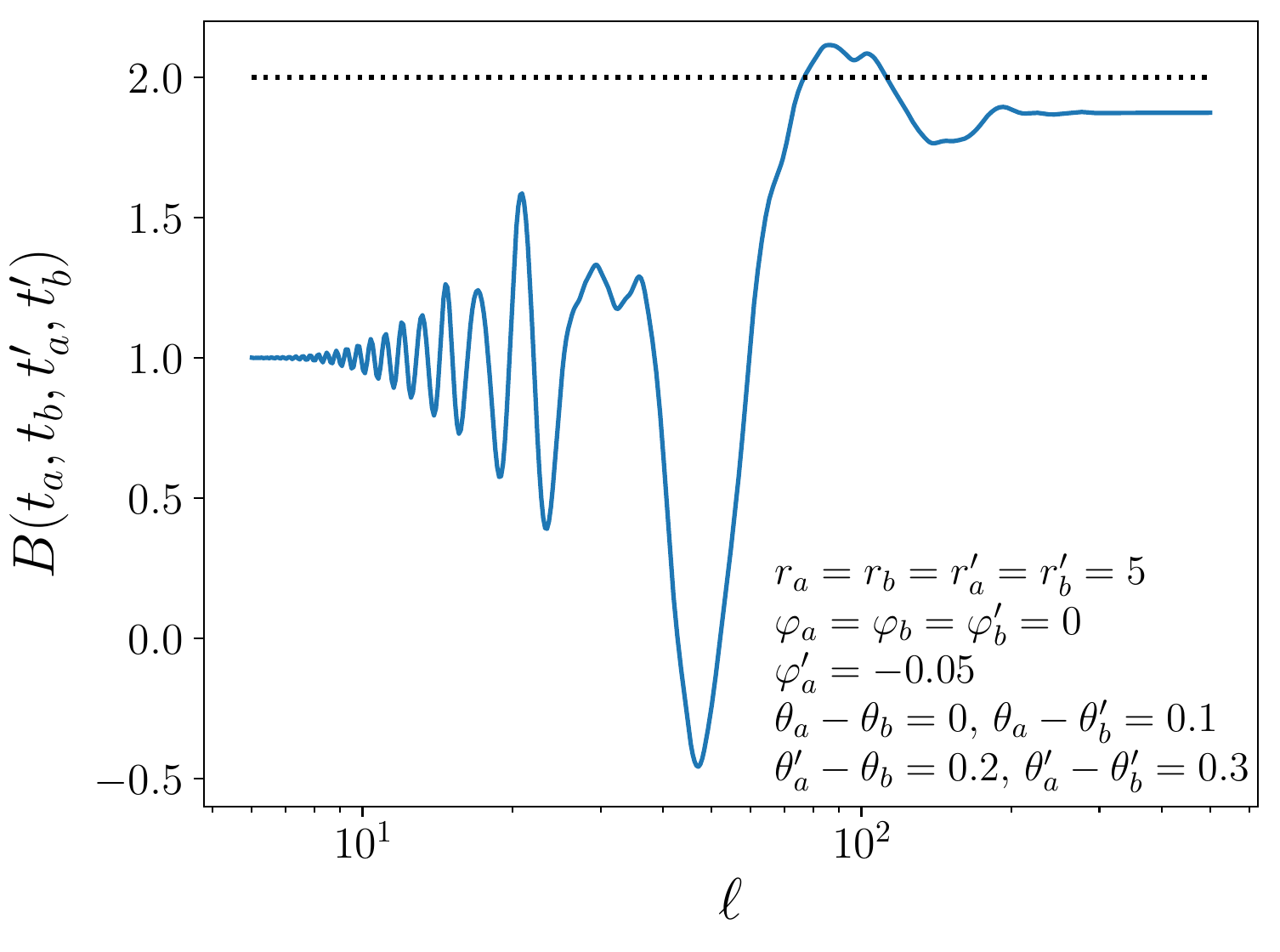}
\caption{Expectation value of the Bell operator $B(t_a,t_b,t_a',t_b')$ as a function of $\ell$, where the parameters specifying the state of the systems at times $t_a$, $t_b$, $t_a'$ and $t_b'$ have been fixed to the values given in the figure. The black dotted line stands for $B = 2$, above which a violation occurs. On can see that the maximal violation is obtained around intermediate values of $\ell$ (namely, according to the above discussion, around $\ell \sim \ee^r$).}  
\label{fig:Bell_to_ell}
\end{center}
\end{figure}

Now that we have made clear how to compute the correlation function $E(t_a,t_b)$, the result can be plugged into \Eq{eq:Bell:inequality} and one can test for violations of bipartite temporal Bell inequalities. As explained at the beginning of \Sec{sec:Analytical:Limits}, $11$ parameters are required to specify the state of the systems at times $t_a$, $t_b$, $t_a'$ and $t_b'$, given that only the changes in the rotation angles matter, and not their individual values. One should also add the spin operator parameter $\ell$, which the observer can in principle set in a free way. This leaves us with $12$ parameters. We have not performed a comprehensive analysis of this whole, high-dimensional parameter space but have instead considered some two-dimensional slices, which is enough to prove that indeed, bipartite temporal Bell inequalities can be violated by two-mode squeezed states. 

Our strategy is that since we are searching for violations of the Bell inequality, $B>2$, it seems reasonable to focus on parameters that make the first correlator appearing in $B$, $E(t_a,t_b)$, close to unity. We already know that $E(t_a,t_b)$ is close to one when $(\varphi_a,\varphi_b) = (-\Delta\theta/2,\Delta\theta/2)$, for a given $\Delta\theta$, in the large squeezing regime (see \Sec{sec:large:ell:large:squeezing:limit}). This is why in the following, we set $\theta_a - \theta_b = 0$ and $\varphi_a = \varphi_b = 0$. In \Fig{fig:Bell_to_ell}, we display the expectation value of the Bell operator as a function of $\ell$, where the other parameters have been fixed according to that strategy. One can see that a violation is obtained when $\ell$ is ``intermediate'' in the sense discussed above, \ie when $\ell \sim \ee^r$. We will therefore focus on such intermediate values of $\ell$ below.

\begin{figure}[t]
\begin{center}
\includegraphics[width=0.469\textwidth]{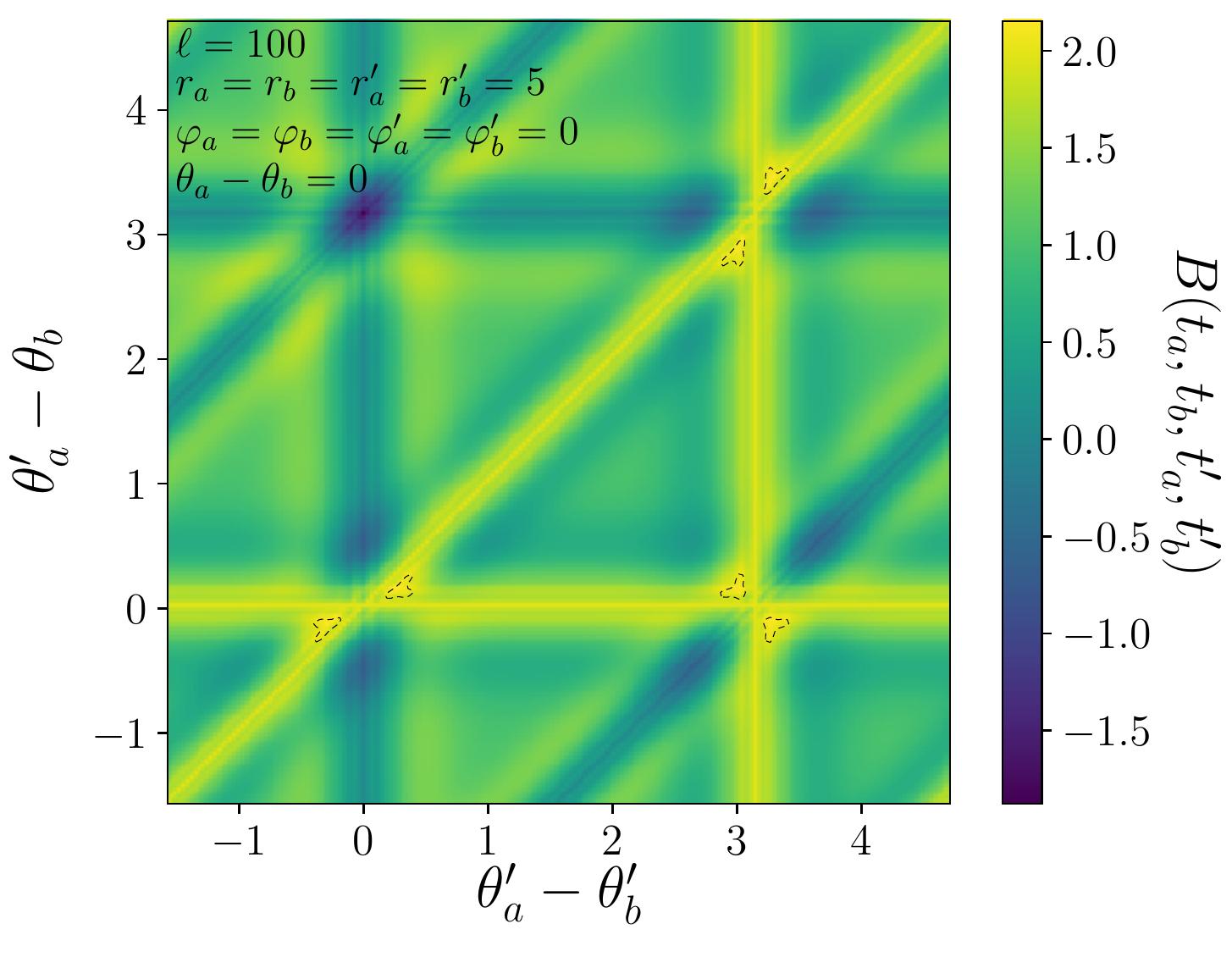}
\includegraphics[width=0.469\textwidth]{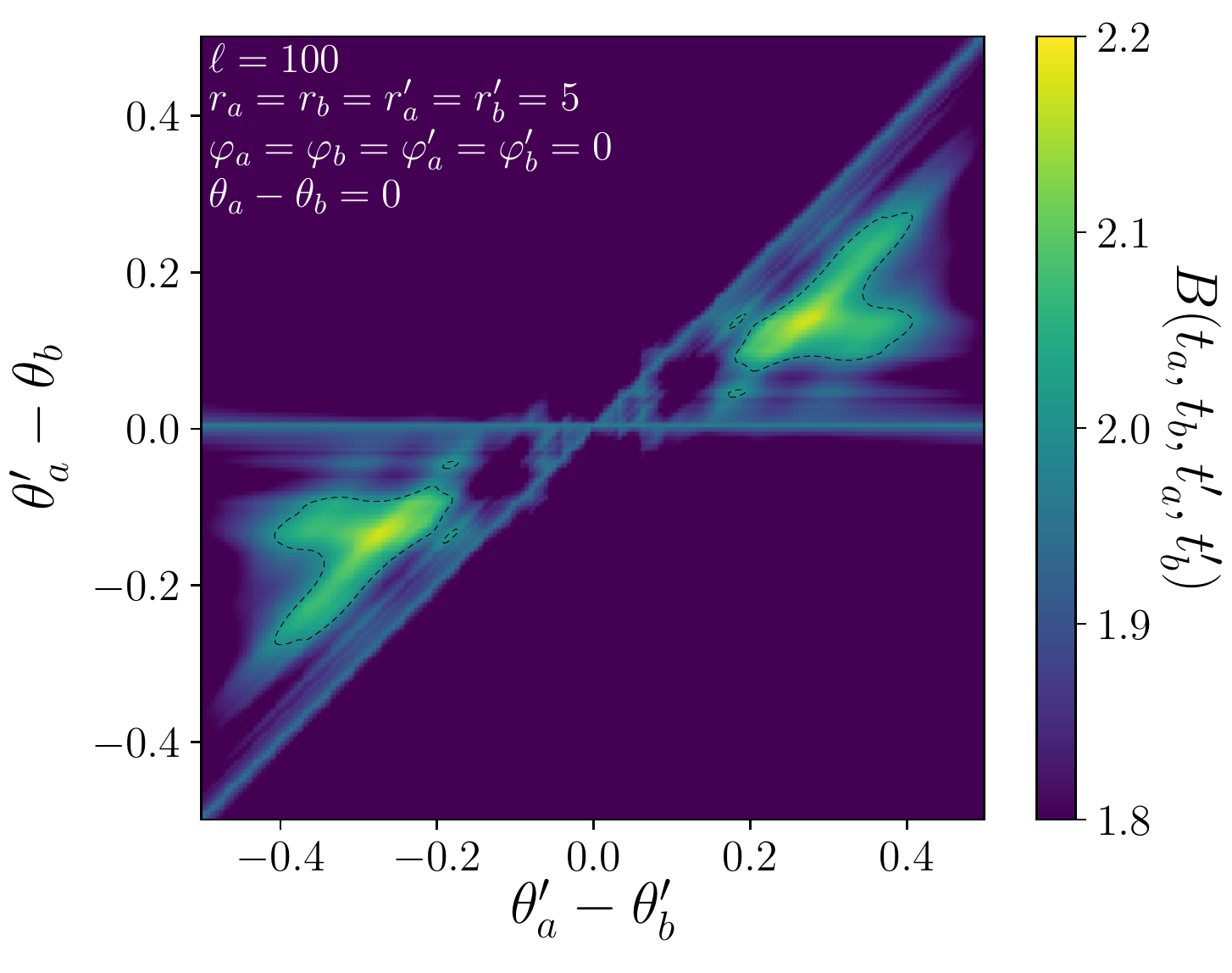}
\caption{Expectation value of the Bell operator $B(t_a,t_b,t_a',t_b')$, for $\ell=100$, $r_a=r_b=r_{a}'=r_{b}'=5$, $\varphi_a=\varphi_b=\varphi_{a'}=\varphi_{b'}=0$ and $\theta_a-\theta_b=0$, as a function of $\theta_a'-\theta_b'$ and $\theta_a'-\theta_b$. The black dashed lines are contours for $B = 2$, so violation occurs inside the contours. In the left panel, the full $2\pi\times2\pi$ parameter space is displayed while the right panel zooms in the region $(\theta_a'-\theta_b',\theta_a'-\theta_b) = (0,0)$ where violation islands exist. The maximum value of $B$ across the entire map is $\simeq 2.18$.}  
\label{fig:Bell_ell100}
\end{center}
\end{figure}

In \Fig{fig:Bell_ell100}, the expectation value of the Bell operator is shown in the case that all squeezing parameters are frozen ($r_a=r_b=r_{a}'=r_{b}'=5$ and $\varphi_a=\varphi_b=\varphi_{a'}=\varphi_{b'}=0$) and only the rotation angles vary. This is because, in experiments where measurements are performed at a single time, the rotation angle only determines an overall, irrelevant phase of the wavefunction. This is why most analyses of the two-mode squeezed states discard it. As argued above, for multiple-time measurements this is not the case anymore, and we would like to determine how important the rotation angle becomes. The black lines in \Fig{fig:Bell_ell100} are contours of $B=2$, above which the bipartite temporal Bell inequality is violated. As one can see, there are islands in parameters space, inside the contours, where the inequality is indeed violated, and the rotation angles play a crucial role in determining whether or not this is the case. These islands are located around the three points $(\theta_a'-\theta_b',\theta_a'-\theta_b) = (0,0),(\pi,0)$ and $(\pi,\pi)$, and the right panel of \Fig{fig:Bell_ell100} zooms in one of these points (the detailed structure of the map is similar at each of these points). At those points exactly, one can check that each correlator $E$ involved in \Eq{eq:Bell:inequality} is close to $\pm 1$, but they cancel each other out in such a way that no violation occurs. One therefore has to move slightly away from those points, and in the right panel of \Fig{fig:Bell_ell100} one can see that smaller, secondary islands actually exist. The structure of the violation map is therefore rather involved.

\begin{figure}[t]
\begin{center}
\includegraphics[width=0.469\textwidth]{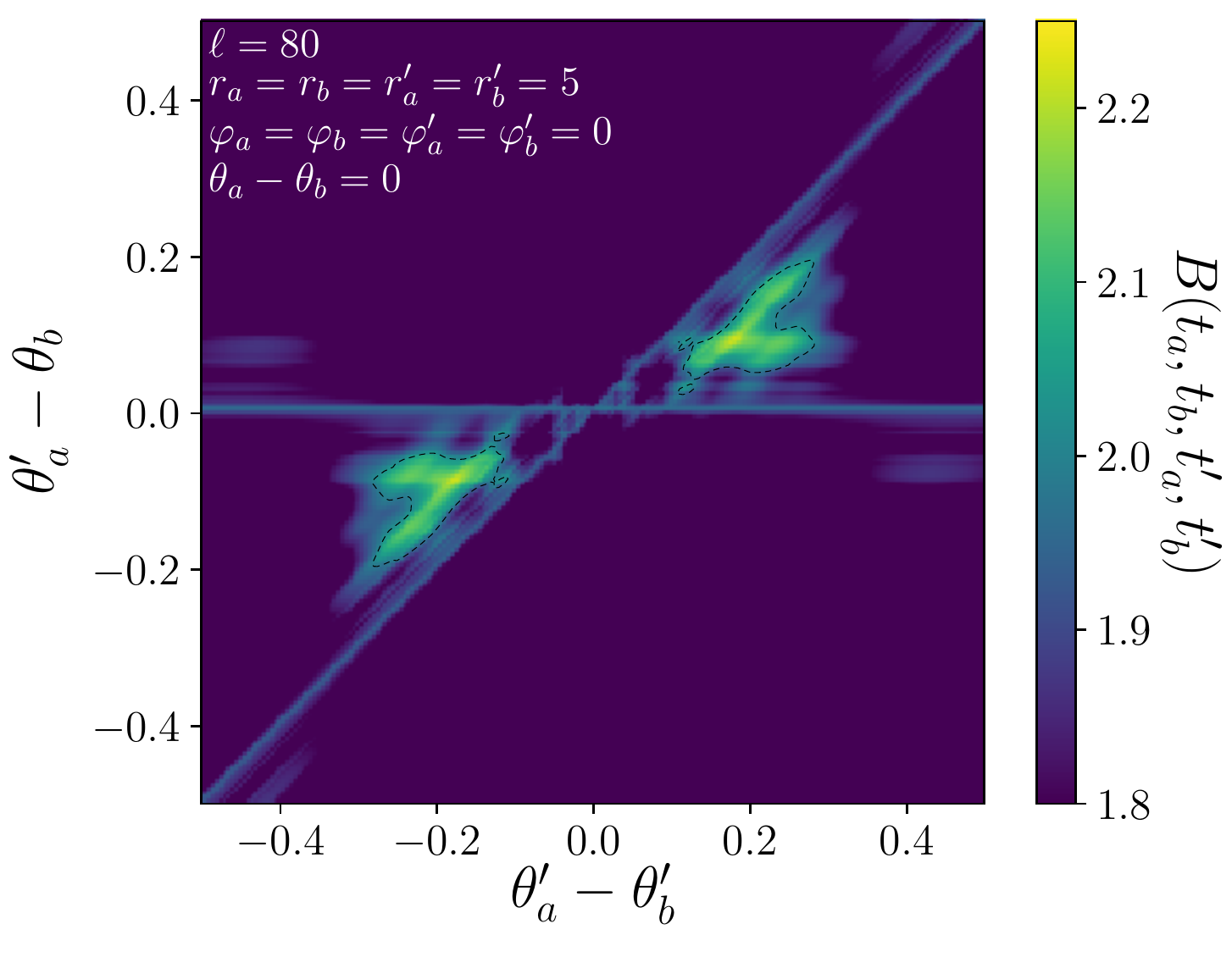}
\includegraphics[width=0.469\textwidth]{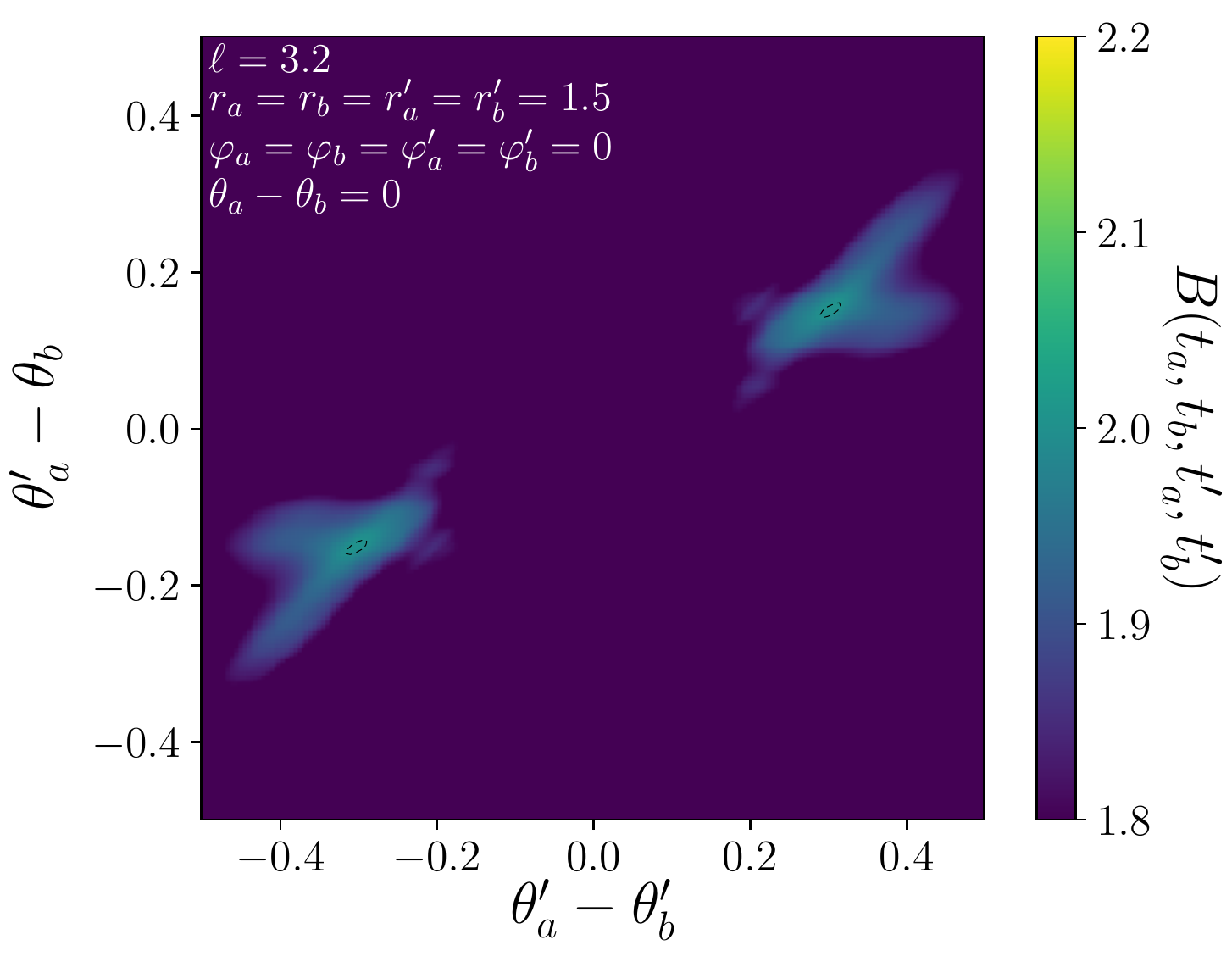}
\caption{Expectation value of the Bell operator $B(t_a,t_b,t_a',t_b')$ as a function of $\theta_a'-\theta_b'$ and $\theta_a'-\theta_b$. Left panel: $\ell=100$, $r_a=r_b=r_{a}'=r_{b}'=5$, $\varphi_a=\varphi_b=\varphi_{a'}=\varphi_{b'}=0$ and $\theta_a-\theta_b=0$. The maximum value of $B$ across the entire map is $\simeq 2.22$. Right panel: $\ell=3.2$ (optimised to get maximal violation), $r_a=r_b=r_{a}'=r_{b}'=1.5$, $\varphi_a=\varphi_b=\varphi_{a'}=\varphi_{b'}=0$, and $\theta_a-\theta_b=0$. The maximum value of $B$ across the entire map is $\simeq 2.00$ up to numerical precision.
In both panels, the black dashed lines are contours for $B = 2$.}
\label{fig:Bell:rotation:angles:more}
\end{center}
\end{figure} 
We have checked that no violation occurs in the infinite-$\ell$ limit. At finite $\ell$, as explained above (see the discussion around the right panel of \Fig{fig:E12:Numerical}), oscillatory features appear in each correlation function $E$, around points where $E\simeq \pm1$ and this leads to the violations. Bipartite temporal Bell inequality violations seem therefore to require $\ell \lesssim e^r$. In order to better see the role played by $\ell$, in the left panel of \Fig{fig:Bell:rotation:angles:more}, the same map as in the right panel of \Fig{fig:Bell_ell100} is displayed but with $\ell =80$. The violation islands still exist. While they are smaller, they are also higher (the maximal value of $B$ in \Fig{fig:Bell_ell100} is found to be $B_\umax \simeq 2.18$ while in the left panel of \Fig{fig:Bell:rotation:angles:more}, it is $B_\umax \simeq 2.22$, and we recall that the Cirel'son bound~\cite{1980LMaPh...4...93C} is given by $B\leq 2\sqrt{2}\simeq 2.83$). We have also tried to decrease the squeezing amplitude starting from the configuration displayed in these figures, allowing us to choose the value of $\ell$ that leads to the maximal violation. In the right panel of \Fig{fig:Bell:rotation:angles:more}, we show the result for $r=1.5$, where the violation is maximal for $\ell\simeq 3.2$. There, the islands have almost disappeared, and the maximal value one obtains is $B_\umax \simeq 2.00$ up to numerical precision. In these slices, Bell inequality violations seem therefore to require $r\gtrsim 1.5$.

\begin{figure}[t]
\begin{center}
\includegraphics[width=0.469\textwidth]{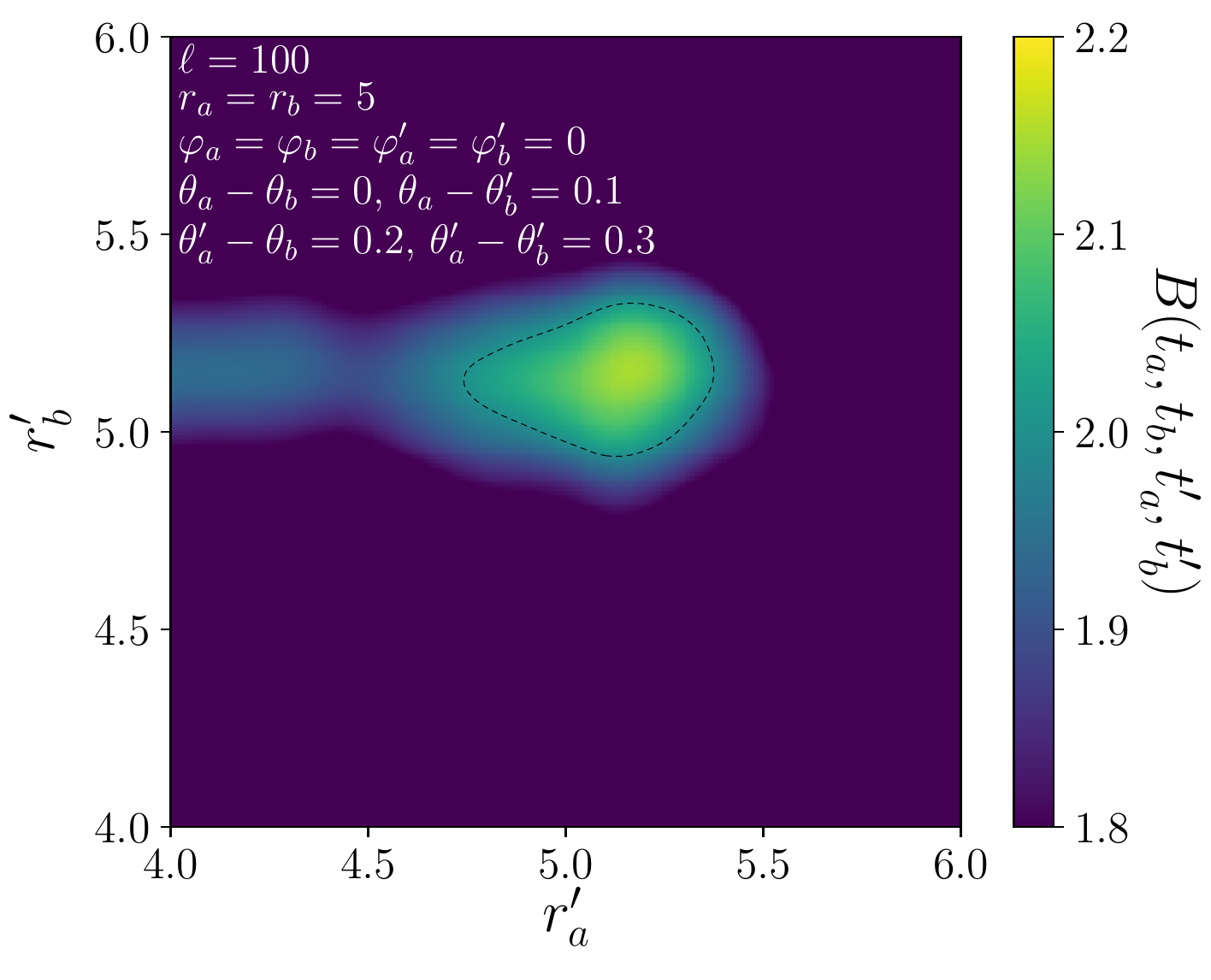}
\includegraphics[width=0.469\textwidth]{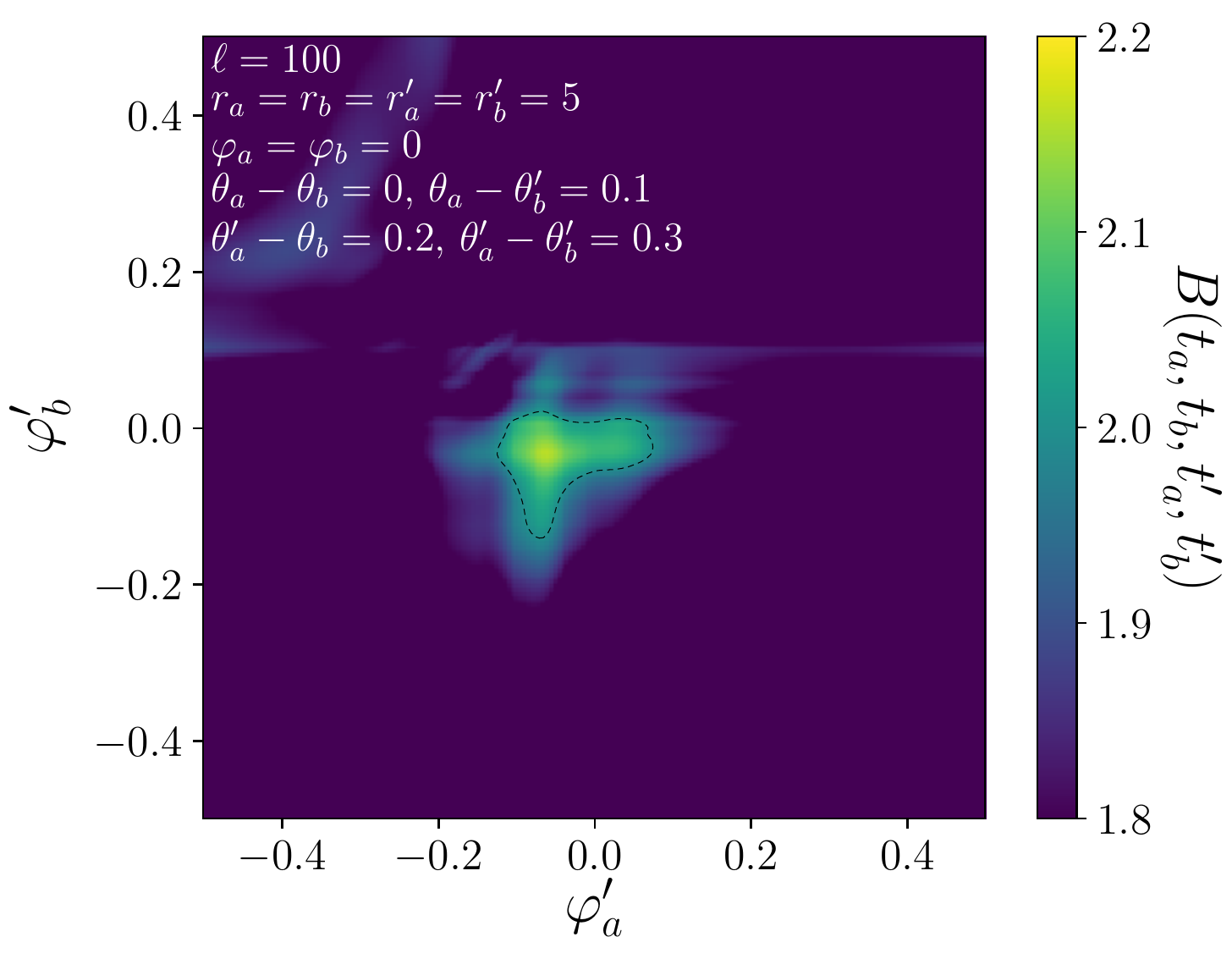}
\caption{Expectation value of the Bell operator $B(t_a,t_b,t_a',t_b')$ as a function of $r_a'$ and $r_b'$ (left panel), and  $\varphi_a'$ and $\varphi_b'$ (right panel), around the violations islands of \Fig{fig:Bell_ell100}.}
\label{fig:Bell:r:varphi}
\end{center}
\end{figure}
Finally, in order to study how the temporal Bell inequality violation depends on the squeezing amplitudes and angles, in \Fig{fig:Bell:r:varphi}, we make the parameters $r_a'$ and $r_b'$ (left panel), and $\varphi_a'$ and $\varphi_b'$ (right panel), vary, starting from close to the maximal violation point of  \Fig{fig:Bell_ell100}. One can see that some amount of fine tuning in these parameters is also necessary to achieve violation.
\section{Conclusion}
\label{sec:conslusion}
In this work we have studied bipartite temporal Bell inequalities with two-mode squeezed states. Such inequalities have the ability to test for realism and locality, while requiring position measurements only. This is particularly useful for experimental setups in which momentum observables cannot be directly accessed, as for instance in the cosmological context. Two-mode squeezed states are continuous, entangled Gaussian states, so we had to introduce a dichotomic, spin-like observable for continuous systems, which we did via \Eq{eq:Si:def}. This operator turns the position into $+1$ or $-1$, depending on which interval of size $\ell$ it falls. We have shown how to compute the bipartite two-point function of that operator, and have studied various analytical limits that were useful to guide our numerical exploration, which is otherwise tedious due to the high dimensionality of the problem. We have then exhibited configurations where the Bell inequality is violated, confirming that two-mode squeezed states have the ability to violate bipartite temporal Bell inequalities. This is clearly the main result of this work.

When $\ell$ is infinite, the pseudo-spin operator becomes the sign operator, which returns $+1$ is the position is positive and $-1$ otherwise, and we could not find configurations leading to successful violation in that case. Optimising the value of $\ell$, which is in principe left to the observer to freely chose,  seems therefore to be crucial. We have also highlighted the role played by the rotation angle. Two-mode squeezed states are characterised by three parameters; the squeezing amplitude, the squeezing angle and the rotation angle; but in single-time measurements, the rotation angle simply sets an overall phase in the wavefunction of the two-mode squeezed state, and is thus irrelevant. It is therefore discarded in most analyses of two-mode squeezed states. However, we have shown that it plays a crucial role in the present context, where measurements are performed at different times, between which the rotation angle is liable to evolve, and the phase difference between the various measurements becomes an important parameter to determine whether or not Bell inequalities are violated. Let us also stress that the dynamics of the rotation angle is set by the Hamiltonian of the system, so probing the part of the Hamiltonian that drives the rotation angle is only possible if multiple-time measurements are performed. This work therefore lays the ground for more thorough investigations of physical systems leading to squeezed states. In fact, in one-mode squeezed states as well, the rotation angle becomes relevant for multiple-time measurements. We plan to investigate this effect, in the context of Leggett-Garg inequalities (for one-mode and two-mode squeezed states) in a future work in preparation. 
We also plan to study how quantum decoherence reduces Bell inequality violations in these contexts.
\section*{Acknowledgements}
It is a pleasure to thank J\'er\^ome Martin for useful discussions and comments on the draft.
K.A. is supported by JSPS Research Fellowships for Young Scientists Grant No. 18J21906, and Advanced Leading Graduate Course for Photon Science.


\appendix

\section{Wavefunction of the two-mode squeezed state}
\label{App:ket:tmss}
In this appendix we provide a derivation of the expression~\eqref{2sq} for the two-mode squeezed state. From the expression of the rotation operator in terms of the number of particles operators, \Eq{eq:R:def}, it is clear that the vacuum state is invariant under rotations, \ie $\hat{R}(t)\left\vert 0,0 \right\rangle = \left\vert 0,0 \right\rangle$. The two-mode squeezed state is thus given by $\Ket{\Psi_{\mathrm{2sq}}(t)} = \hat U_{\mathrm{S}}(t) \hat{R}(t) \Ket{0,0} =  \hat U_{\mathrm{S}}(t) \Ket{0,0}$, where the squeezing operator is given by \Eq{eq:US:def}. We rewrite this expression as
\bea
\label{eq:US:def:alpha}
	\hat U_{\mathrm{S}}(t) = \ee^{\alpha^* \hat{A}^\dagger - \alpha \hat{A} }\, ,
\eea
with $\alpha=r \ee^{2i\varphi}$ and $\hat{A} =\hat c_1\hat c_2 $. The idea is to make use of operator ordering theorems to rewrite \Eq{eq:US:def:alpha} as a product of exponentials that can be easily applied onto the vacuum state, following similar lines as those presented in section 3.3 of \Refa{BarnettRadmore}. 

Our first step is to study the algebra generated by the operators appearing in \Eq{eq:US:def:alpha}, $\hat{A}$ and $\hat{A}^\dagger$. Introducing the Hermitian operator $\hat{B}\equiv \hat{c}_1 \hat{c}_1^\dagger + \hat{c}_2^\dagger \hat{c}_2 = \hat{B}^\dagger$, one can first check that $\hat{A}$, $\hat{A}^\dagger$ and $B$ form a closed algebra, with
\bea
\label{eq:com:A_Adag}
\left[\hat{A},\hat{A}^\dag\right]=B\, ,\quad\quad
\left[\hat{A},\hat{B}\right]=2 \hat{A}\, ,\quad\quad
\left[\hat{A}^\dagger,\hat{B}\right]=-2 \hat{A}^\dagger\, .
\eea
All commutators within this algebra can be computed with these formulas using iterative methods. In particular, one finds
\bea
\left[\hat{A}^n,\hat{B}\right]&=&2n\hat{A}^n\\
\label{eq:com:Adag_n:B}
\left[\left(\hat{A}^\dagger\right)^n,\hat{B}\right]&=&-2n\left(\hat{A}^\dagger\right)^n\\
\left[\hat{A}^n,\hat{A}^\dagger\right]&=&n\hat{A}^{n-1}\hat{B}-n(n-1)\hat{A}^{n-1}\\
\label{eq:com:Adag_n:A}
\left[ \left(\hat{A}^\dagger\right)^n ,\hat{A}\right] &=& -n \hat{B}\left(\hat{A}^\dagger\right)^{n-1}+n\left(n-1\right)\left(\hat{A}^\dagger\right)^{n-1}\\
\label{eq:com:B_n:A}
\left[\hat{A},\hat{B}^n\right]&=& \hat{A}\left[ \hat{B}^n-\left(\hat{B}-2\right)^n\right] = \left[\left(\hat{B}+2\right)^n-\hat{B}^n\right]\hat{A}\\
\left[\hat{A}^\dagger,\hat{B}^n\right]&=& \hat{A}^\dagger \left[ \hat{B}^n-\left(\hat{B}+2\right)^n\right] = \left[\left(\hat{B}-2\right)^n-\hat{B}^n\right]\hat{A}^\dagger
\eea
for any integer number $n$. From here, commutators involving exponentials can be readily derived. Making use of \Eqs{eq:com:Adag_n:B}, \eqref{eq:com:B_n:A} and~\eqref{eq:com:Adag_n:A}, one respectively finds three formulas that will turn out to be useful below, namely
\bea
\label{eq:com:exp:Adag:B}
\left[\ee^{z \hat{A}^\dagger},\hat{B}\right]&=&-2z\hat{A}^\dagger \ee^{z \hat{A}^\dagger}\\
\label{eq:com:expB:hatA}
\left[\ee^{z\hat{B}},\hat{A}\right]&=&\left(\ee^{-2z}-1\right)\hat{A}\ee^{z\hat{B}}\\
\label{eq:com:expAdag:A}
\left[\ee^{z \hat{A}^\dagger},\hat{A}\right]&=&\left(z^2\hat{A}^\dagger-z\hat{B}\right)\ee^{z\hat{A}^\dagger}
\eea
for any complex number $z$.

Our second step is to introduce the function
\bea
\label{eq:F(x):1}
F(x)=\ee^{x\left(\alpha^* \hat{A}^\dagger - \alpha \hat{A}\right)}\, ,
\eea
such that $\hat U_{\mathrm{S}} = F(1)$, and to study that function. The goal is to rewrite $F(x)$ as a product of exponentials. These exponentials must involve elements of the algebra only, which allow us to introduce the ansatz
\bea
\label{eq:F(x):2}
F(x)=\ee^{f(x) \hat{A}^\dagger} \ee^{g(x) \hat{B}} \ee^{h(x) \hat{A}}
\eea
where $f$, $g$ and $h$ are three functions to determine. This can be done by differentiating $F$ with respect to $x$. Making use of \Eq{eq:F(x):1}, one has
\bea
\label{eq:F'(x):1}
F'(x)=\left(\alpha^* \hat{A}^\dagger - \alpha \hat{A}\right) F(x)\, ,
\eea
while \Eq{eq:F(x):2} gives rise to three terms, namely
\bea
F'(x) &=& f'(x) \hat{A}^\dagger \ee^{f(x) \hat{A}^\dagger} \ee^{g(x) \hat{B}} \ee^{h(x) \hat{A}}
 + g'(x) \ee^{f(x) \hat{A}^\dagger} \hat{B} \ee^{g(x) \hat{B}} \ee^{h(x) \hat{A}}
 + h'(x) \ee^{f(x) \hat{A}^\dagger} \ee^{g(x) \hat{B}} \hat{A} \ee^{h(x) \hat{A}}\, .
 \nonumber \\ & &
\eea
In this expression, the first term is simply given by $ f'(x) \hat{A}^\dagger F(x) $. For the second term, making use of \Eq{eq:com:exp:Adag:B} to rewrite $\ee^{f(x) \hat{A}^\dagger} \hat{B} =  \hat{B} \ee^{f(x) \hat{A}^\dagger} -2 f(x) \hat{A}^\dagger \ee^{f(x) \hat{A}^\dagger}$, one finds that it is given by $g'(x)[\hat{B}-2f(x)\hat{A}^\dagger]F(x)$. The third term can be computed similarly by first making use of \Eq{eq:com:expB:hatA} and then of \Eq{eq:com:expAdag:A}, and one obtains $h'(x)\ee^{-2g(x)} [f^2(x)\hat{A}^\dag +\hat{A}-f(x)\hat{B}] F(x)$. Combining these results together, one obtains
\bea
\label{eq:F'(x):2}
F'(x) &=& \left\lbrace h'(x)\ee^{-2g(x)} \hat{A} +\left[f'(x)-2f(x)g'(x)+h'(x)\ee^{-2g(x)} f^2(x)\right]\hat{A}^\dagger
\right. \nonumber \\ & &\left.
+\left[g'(x)-h'(x)f(x)\ee^{-2g(x)}\right]\hat{B} \right\rbrace F(x)\, .
\eea
By identifying \Eqs{eq:F'(x):1} and~\eqref{eq:F'(x):2}, one obtains three coupled differential equations for the functions $f$, $g$ and $h$, namely
\bea
\label{eq:equadiff:1}
f'-2f g'+h' f^2 \ee^{-2g} =\alpha^*\, ,\\
\label{eq:equadiff:2}
h'\ee^{-2g}=-\alpha\, ,\\
\label{eq:equadiff:3}
g'-fh' \ee^{-2g}=0\, .
\eea
This system must be solved with the boundary conditions $f(0)=g(0)=h(0)=0$ that simply follow from identifying \Eqs{eq:F(x):1} and~\eqref{eq:F(x):2} when $x=0$. This can be done as follows. Plugging \Eq{eq:equadiff:3} into \Eq{eq:equadiff:2}, one obtains $g'=-\alpha f$, and plugging that relation together with \Eq{eq:equadiff:3} into \Eq{eq:equadiff:1} gives rise to $f'+\alpha f^2 = \alpha^*$. This can be readily integrated, and imposing that $f(0)=0$, one obtains
\bea
f(x)=\sqrt{\frac{\alpha^*}{\alpha}}\tanh\left(\vert \alpha\vert x\right)\, .
\eea
From here, the relation $g'=-\alpha f$ can also be integrated, and imposing that $g(0)=0$ leads to
\bea
g(x)=-\ln\left[\cosh\left(\vert \alpha\vert x\right)\right]\, .
\eea
Finally, \Eq{eq:equadiff:3} gives rise to $h'=-\alpha \ee^{2g}$, which can be integrated as
\bea
h(x)=-\sqrt{\frac{\alpha}{\alpha^*}}\tanh\left(\vert \alpha\vert x\right)
\eea
where we have used that $h(0)=0$. Evaluating the three functions $f$, $g$ and $h$ when $x=1$, and expressing $\alpha$ in terms of $r$ and $\varphi$, one obtains for the squeezing operator
\bea
\label{eq:US:ordered}
\hat U_{\mathrm{S}} = 
\exp\left[\ee^{-2i\varphi} \tanh(r) \hat{A}^\dagger \right]
\exp\left[-\ln \left(\cosh r\right) \hat{B} \right]
\exp\left[-\ee^{2i\varphi} \tanh(r) \hat{A} \right] ,
\eea
which is the operator ordered expression we were seeking.

We can now apply the squeezing operator onto the vacuum state. Recalling that $\hat{A} =\hat c_1\hat c_2 $ and $\hat{B}\equiv \hat{c}_1 \hat{c}_1^\dagger + \hat{c}_2^\dagger \hat{c}_2 $, one can see that $\hat{A}$ annihilates the vacuum state, while $\hat{B}$ leaves it invariant, $\hat{B}\Ket{0,0}=\Ket{0,0}$. When applied to the vacuum state, the last exponential terms in \Eq{eq:US:ordered} has therefore no effect, while the second terms adds a prefactor $1/\cosh(r)$. Taylor expanding the first exponential term, one then obtains
\bea
\label{eq:tmss:n:app}
\Ket{\Psi_{\mathrm{2sq}}}=  \frac{1}{\cosh(r)} \sum_{n=0}^{\infty} \ee^{-2i n \varphi}\tanh^n(r)\Ket{n,n}\, ,
\eea 
which coincides with \Eq{2sq} given in the main text.

This two-mode squeezed state can finally be expressed in position space, and the wavefunction is given by
\bea
\label{eq:2mss:position:1}
\Psi_{\mathrm{2sq}}\left(Q_1, Q_2\right) = \left\langle Q_1,Q_2\right\vert\left.\Psi_{\mathrm{2sq}}\right\rangle=
\frac{1}{\cosh(r)} \sum_{n=0}^{\infty} \ee^{-2i n \varphi}\tanh^n(r) 
\left\langle Q_1\right\vert\left. n \right\rangle\left\langle Q_2\right\vert\left. n \right\rangle\, .
\eea 
The scalar products $\left\langle Q\right\vert\left. n \right\rangle$ can be expressed in terms of the Hermite potentials $H_n(Q)$, \ie
\bea
\left\langle Q\right\vert\left. n \right\rangle = \frac{\pi^{-1/4}}{\sqrt{2^n n!}}\ee^{-\frac{Q^2}{2}} H_n(Q)\, .
\eea
The sum over $n$ appearing in \Eq{eq:2mss:position:1} can then be performed by means of Eq.~(18.18.28) of \Refa{2011ConPh..52..497T}, namely\footnote{Hereafter, unless specified otherwise, the square root of a complex number $Z \notin \mathbb{R}^-$ is given by $\sqrt{Z}=\sqrt{\rho}\ee^{i\gamma/2}$, for $Z=\rho \ee^{i \gamma}$ with $-\pi<\gamma<\pi$.
Notice that, for the square roots appearing in \Eqs{Muler} and \eqref{2sq_wave_func}, $\gamma$ actually lies within $[-\pi/2, \pi/2]$.
\label{footnote:sqrt:complex}
}
\bea
\sum_{n=0}^\infty \frac{H_n(x) H_n(y)}{2^n n!}z^n = \left(1-z^2\right)^{-1/2}\exp\left[\frac{2xyz-\left(x^2+y^2\right)z^2}{1-z^2}\right]\, ,
\label{Muler}
\eea
with $\abs{z}<1$.
This gives rise to
\bea	
\Psi_\mathrm{2sq}( Q_1, Q_2)
	=\frac{\exp\com{\frac{1}{2}A(r,\varphi)\prn{Q_1^2 + Q_2^2} + B(r,\varphi)Q_1Q_2 }}{\cosh r\,\sqrt \pi \sqrt{1 - e^{-4i\varphi}\tanh^2 r}} 
	\label{2sq_wave_func}
\eea
where the functions $A(r,\varphi)$ and $B(r, \varphi)$ are given by
\bea
	A(r,\varphi) = -\frac{1+e^{-4i\varphi}\tanh^2r}{1-e^{-4i\varphi}\tanh^2r}
	,\quad
	B(r,\varphi) = \frac{2e^{-2i\varphi}\tanh r}{1-e^{-4i\varphi}\tanh^2r}\, .
	\label{AB}
\eea

\section{Correlation function of the evolution operator}
\label{app:correlation}
In this appendix, we compute the two-point function of the evolution operator appearing in \Eq{Eab_detail}, namely $\Braket{\tilde Q_1, \tilde Q_2|\hat U(t_a) \hat U^\dagger(t_b)| \bar Q_1, \bar Q_2} $. We first introduce the two-mode coherent states $\ket{\bm{u}}=\ket{u_1,u_2}$, which are eigenstates of the annihilation operators 
\bea
\label{eq:coherent:state:def}
\hat{c}_i \ket{\bm{u}} = u_i \ket{\bm{u}}
\eea
where $i=1$ or $2$. Decomposing the eigenvalues into real and imaginary parts,
\bea
\label{eq:u:R:I}
	u_i=\frac{u_{i\RR} + iu_{i\II}}{\sqrt 2},\,
\eea
and introducing the integration element $\dd\bm{u} = \dd u_{1\RR}\dd u_{1\II} \dd u_{2\RR}\dd u_{2\II}$, they satisfy the closure relation
\bea
\int\frac{\dd \bm{u}}{\left(2\pi\right)^2}\Ket{\bm{u}}\Bra{\bm{u}}=1\, .
\eea
One can plug this closure relation on each side of the evolution operators in the two-point correlator we aim at computing, leading to
\bea
    \hspace{-5mm}
	\Braket{\tilde Q_1, \tilde Q_2|\hat U(t_a) \hat U^\dagger(t_b)| \bar Q_1, \bar Q_2} 
	&=& \int\frac{\dd \bvec u}{(2\pi)^2}\int\frac{\dd \bvec v}{(2\pi)^2}\int\frac{\dd \bvec w}{(2\pi)^2}
	\Braket{\tilde Q_1 | w_1}\Braket{\tilde Q_2 | w_2}
	\nonumber \\
	 && \Braket{\bvec w | \hat U(t_a) | \bvec u}
	\Braket{\bvec u | \hat U^\dagger(t_b) | \bvec v} \Braket{v_1 | \bar Q_1}\Braket{v_2 | \bar Q_2}.
\eea
In this expression, $\Braket{\tilde Q_1 | w_1}$ is the wave function of the coherent state, and is given by
\bea
\label{eq:ScalarProduct:Q:w}
	\Braket{\tilde Q_1 | w_1} 
	&=& \frac{1}{\pi^{1/4}}\exp\com{-\frac{i}{2}w_{1\RR}w_{1\II} + iw_{1\II}\tilde Q_1 -\frac{1}{2}(\tilde Q_1 - w_{1\RR})^2   } \, ,
\eea
with a similar expression for $\Braket{\tilde Q_2 | w_2}$, $\Braket{v_1 | \bar Q_1}$ and $\Braket{v_2 | \bar Q_2}$. Let us then consider $\Braket{\bvec w | \hat U(t_a) | \bvec u} = \Braket{\bvec w | \hat U_\mathrm{S}(t_a) \hat R(t_a) | \bvec u}$. By using the decomposition
\bea
\label{eq:coherent:state:number:of:particules}
\Ket{\bm{u}} = \ee^{-\frac{\left\vert u_1\right\vert^2+\left\vert u_2\right\vert^2}{2}}\sum_{n_1=0}^\infty \sum_{n_2=0}^\infty \frac{u_1^{n_1}}{\sqrt{n_1}!}\frac{u_2^{n_2}}{\sqrt{n_2!}} \Ket{n_1,n_2},
\eea
\Eq{eq:R:def} gives rise to 
\bea
\hat{R} \Ket{\bm{u}} &=& 
\sum_{m_1=0}^\infty\sum_{m_2=0}^\infty \ee^{-\frac{\left\vert u_1\right\vert^2+\left\vert u_2\right\vert^2}{2}}\sum_{n_1=0}^\infty \sum_{n_2=0}^\infty 
\frac{u_1^{n_1}}{\sqrt{n_1!}} \left(i\theta n_1\right)^{m_1}
\frac{u_2^{n_2}}{\sqrt{n_2!}} \left(i\theta n_2\right)^{m_2}
\Ket{n_1,n_2}\\ &=&
\ee^{-\frac{\left\vert u_1\right\vert^2+\left\vert u_2\right\vert^2}{2}}\sum_{n_1=0}^\infty \sum_{n_2=0}^\infty 
\frac{u_1^{n_1}}{\sqrt{n_1!}} \ee^{i\theta n_1}
\frac{u_2^{n_2}}{\sqrt{n_2!}} \ee^{i\theta n_2}
\Ket{n_1,n_2}\\ &=&
\ee^{-\frac{\left\vert u_1\right\vert^2+\left\vert u_2\right\vert^2}{2}}\sum_{n_1=0}^\infty \sum_{n_2=0}^\infty 
\frac{\left(u_1\ee^{i\theta}\right)^{n_1}}{\sqrt{n_1!}} 
\frac{\left(u_2\ee^{i\theta}\right)^{n_2}}{\sqrt{n_2!}} 
\Ket{n_1,n_2}\\ &=& \Ket{e^{i\theta_a}u_1,e^{i\theta_a}u_2} = \Ket{e^{i\theta}\bm{u}} .
\eea
One thus has $\Braket{\bvec w | \hat U(t_a) | \bvec u} = \Braket{\bvec w | \hat U_\mathrm{S}(t_a)  | \ee^{i \theta_a} \bvec u}$, where $\theta_a$ is a short-hand notation for $\theta(t_a)$. The next step is to use the operator ordered expression~\eqref{eq:US:ordered} for $\hat{U}_\mathrm{S}$. Recalling that $\hat{A}=\hat{c}_1 \hat{c}_2$ and $\hat{B}=\hat{c}_1 \hat{c}_1^\dagger+\hat{c}_2^\dagger \hat{c}_2=1+\hat{n}_1+\hat{n}_2$, \Eq{eq:coherent:state:def} gives rise to $\hat{A}\Ket{\bm{u}}=u_1 u_2 \Ket{\bm{u}}$, hence $\ee^{z\hat{A}}\Ket{\bm{u}} = \ee^{z u_1 u_2}\Ket{\bm{u}}$ for any complex number $z$. Similarly, one has $\Bra{w}\ee^{z\hat{A}^\dagger} = \ee^{z w_1^* w_2^*}\Bra{w}$. 

Making use of \Eq{eq:coherent:state:number:of:particules}, one also has
\bea
\left\langle \bm{w} \right\vert \ee^{z \hat{n}_1} \left\vert \bm{u} \right\rangle &=& 
\ee^{-\frac{\left\vert u_1\right\vert^2+\left\vert u_2\right\vert^2+\left\vert v_1\right\vert^2+\left\vert v_2\right\vert^2}{2}}
\sum_{n,m,k} \frac{z^k n^k}{k!} \frac{\left(w_1^* u_1 \right)^n}{n!} \frac{\left(w_2^* u_2 \right)^m}{m!} 
\\ & = &
\ee^{-\frac{\left\vert u_1\right\vert^2+\left\vert u_2\right\vert^2+\left\vert v_1\right\vert^2+\left\vert v_2\right\vert^2}{2}}
\sum_{n,m} \frac{\left(w_1^* u_1 \ee^z \right)^n}{n!} \frac{\left(w_2^* u_2 \right)^m}{m!} 
\\ & = &
\ee^{-\frac{\left\vert u_1\right\vert^2+\left\vert u_2\right\vert^2+\left\vert v_1\right\vert^2+\left\vert v_2\right\vert^2}{2}+w_1^* u_1 \ee^z+w_2^* u_2}
\eea
for any complex number $z$, and a similar expression for $\left\langle \bm{w} \right\vert \ee^{z \hat{n}_1} \left\vert \bm{u} \right\rangle$. Similarly, one finds
\bea
\left\langle \bm{w} \right\vert \ee^{z \hat{B}} \left\vert \bm{u} \right\rangle &=& \ee^{-\frac{\left\vert u_1\right\vert^2+\left\vert u_2\right\vert^2+\left\vert v_1\right\vert^2+\left\vert v_2\right\vert^2}{2}+w_1^* u_1 \ee^z+w_2^* u_2\ee^z +z}.
\eea
Combining the previous results, one obtains
\bea
	\Braket{\bvec w | \hat U(t_a) | \bvec u}
	=\frac{1}{\cosh r_a}\exp(\mathcal F),
\eea
where
\bea
	\mathcal F&=& e^{-2i\varphi_a}\tanh r_a w_1^*w_2^*
	- e^{2i\varphi_a}e^{2i\theta_a}\tanh r_a u_1u_2
	\nonumber \\
	&& - \frac{1}{2}\prn{\abs{w_1}^2 + \abs{u_1}^2 + \abs{w_2}^2 + \abs{u_2}^2 } 
	+\frac{e^{i\theta_a}}{\cosh r_a}\prn{w_1^*u_1 + w_2^*u_2} \\
	&=& \frac{1}{2}e^{-2i\varphi_a}\tanh r_a\prn{w_{1\RR}w_{2\RR} - w_{1\II}w_{2\II} -iw_{1\RR}w_{2\II} -iw_{1\II}w_{2\RR}    } \nonumber \\
	&& -\frac{1}{2}e^{2i\varphi_a}e^{2i\theta_a}\tanh r_a\prn{u_{1\RR}u_{2\RR} - u_{1\II}u_{2\II} +iu_{1\RR}u_{2\II} +iu_{1\II}u_{2\RR}    } \nonumber \\
	&& -\frac{1}{4}\prn{w_{1\RR}^2 + w_{1\II}^2 + w_{2\RR}^2 + w_{2\II}^2 + u_{1\RR}^2 + u_{1\II}^2 + u_{2\RR}^2 + u_{2\II}^2 } \nonumber \\
	&& +\frac{1}{2}\frac{e^{i\theta_a}}{\cosh r_a}
	(w_{1\RR}u_{1\RR} + w_{1\II}u_{1\II} +iw_{1\RR}u_{1\II} -iw_{1\II}u_{1\RR} \nonumber \\
	&&\qquad \qquad \qquad+ w_{2\RR}u_{2\RR} + w_{2\II}u_{2\II} +iw_{2\RR}u_{2\II} -iw_{2\II}u_{2\RR}    ).
\eea
Here in the second expression, we have expanded $u_1$, $u_2$, $w_1$ and $w_2$ into their real and imaginary parts, see \Eq{eq:u:R:I}. The form $\mathcal{F}$ is quadratic in these variables, and since the argument of the exponential in \Eq{eq:ScalarProduct:Q:w} is also quadratic, our result for $\Braket{\tilde Q_1, \tilde Q_2|\hat U(t_a) \hat U^\dagger(t_b)| \bar Q_1, \bar Q_2} $ can be written in matricial form if one introduces the 12-dimensional vector,
\bea
	\alpha^\mathrm{T}\equiv [u_{1\RR}, u_{1\II}, u_{2\RR}, u_{2\II},  v_{1\RR}, v_{1\II}, v_{2\RR}, v_{2\II},  w_{1\RR}, w_{1\II}, w_{2\RR}, w_{2\II} ],
\eea
in terms of which
\bea
	\Braket{\tilde Q_1, \tilde Q_2|\hat U(t_a) \hat U^\dagger(t_b)| \bar Q_1, \bar Q_2}  =
	\frac{1}{64\pi^7}\frac{e^{-\prn{\tilde Q_1^2 + \tilde Q_2^2 + \bar Q_1^2 + \bar Q_2^2}/2}}{\cosh r_a \cosh r_b} 
	\int \dd^{12}\alpha e^{- \frac{1}{2}\alpha^\mathrm{T} M \alpha  - J^\mathrm{T}\alpha},\quad
\eea
where 
\bea
\label{eq:J:def}
J^\mathrm{T} = \com{0,0,0,0,-\bar Q_1,i\bar Q_1,-\bar Q_2,i\bar Q_2,-\tilde Q_1,-i\tilde Q_1,-\tilde Q_2,-i\tilde Q_2}
\eea
and
\begin{eqnarray}
	\hspace{-5mm}
	M=\com{
	\begin{array}{cccccccccccc}
		1 & 0 & \Theta_a + \Theta_b^* & i\Theta_a - i\Theta_b^* & -C_b^* & -iC_b^* & 0 & 0 & -C_a & iC_a & 0 & 0
		\vspace{1mm}\\
		 0 & 1 & i\Theta_a - i\Theta_b^* & -\Theta_a - \Theta_b^* & iC_b^* & -C_b^* & 0 & 0 & -iC_a & -C_a & 0 & 0
		\vspace{1mm}\\
		 \Theta_a + \Theta_b^* & i\Theta_a - i\Theta_b^* & 1 & 0 & 0 & 0 & -C_b^* & -iC_b^* & 0 & 0 & -C_a & iC_a
		\vspace{1mm}\\
		 i\Theta_a - i\Theta_b^* & -\Theta_a - \Theta_b^* & 0 & 1 & 0 & 0 & iC_b^* & -C_b^* & 0 & 0 & -iC_a & -C_a
		\vspace{1mm}\\
		 -C_b^* & iC_b^* & 0 & 0 & \frac{3}{2} & -\frac{i}{2} & -T_b & -iT_b & 0 & 0 & 0 & 0
		\vspace{1mm}\\
		 -iC_b^* & -C_b^* & 0 & 0 & -\frac{i}{2} & \frac{1}{2} & -iT_b & T_b & 0 & 0 & 0 & 0
		\vspace{1mm}\\
		 0 & 0 & -C_b^* & iC_b^* & -T_b & -iT_b & \frac{3}{2} & -\frac{i}{2} & 0 & 0 & 0 & 0
		\vspace{1mm}\\
		 0 & 0 & -iC_b^* & -C_b^* & -iT_b & T_b & -\frac{i}{2} & \frac{1}{2} & 0 & 0 & 0 & 0
		\vspace{1mm}\\
		 -C_a & -iC_a & 0 & 0 & 0 & 0 & 0 & 0 & \frac{3}{2} & \frac{i}{2} & -T_a^* & iT_a^*
		\vspace{1mm}\\
		 iC_a & -C_a & 0 & 0 & 0 & 0 & 0 & 0 & \frac{i}{2} & \frac{1}{2} & iT_a^* & T_a^*
		\vspace{1mm}\\
		 0 & 0 & -C_a & -iC_a & 0 & 0 & 0 & 0 & -T_a^* & iT_a^* & \frac{3}{2} & \frac{i}{2}
		\vspace{1mm}\\
		 0 & 0 & iC_a & -C_a & 0 & 0 & 0 & 0 & iT_a^* & T_a^* & \frac{i}{2} & \frac{1}{2}
	\end{array}
	}
	.
	\nonumber \\ \label{M}
\end{eqnarray}
Here we have introduced $C_a = e^{i\theta_a}/(2\cosh r_a)$, $T_a = (1/2)e^{2i\varphi_a}\tanh r_a$ and $\Theta_a = e^{2i\theta_a}T_a$. If $\det M\neq0$, the Gaussian integral can be performed, and one obtains\footnote{The precise meaning of $\sqrt{\det M}$ is a priori not obvious since $\det M$ is a complex number, and the branch cut of the complex square root function leaves the sign of $\sqrt{\det M}$ ambiguous. However, from \Eq{M}, one can show that $M$ is a symmetric normal matrix, \ie $MM^\dagger = M^\dagger M$. This implies that the real part and the imaginary part of $M$ commute, so they can be simultaneously diagonalised by an orthogonal matrix. The square root of $\det M$ thus stands for the product of the square roots of each eigenvalue of $M$. Since the square root of each eigenvalue is well-defined because all eigenvalues have a positive real part (otherwise the Gaussian integral could not be performed), this removes the ambiguity.
\label{foot_sqrt_detM}
}
\bea
	\Braket{\tilde Q_1, \tilde Q_2|\hat U(t_a) \hat U^\dagger(t_b)| \bar Q_1, \bar Q_2} =
	\frac{1}{\pi}\frac{1}{\cosh r_a \cosh r_b} \frac{1}{\sqrt{\det M}}
	e^{-\frac{1}{2}\prn{\tilde Q_1^2 + \tilde Q_2^2 + \bar Q_1^2 + \bar Q_2^2}+ \frac{1}{2}J^\mathrm{T}M^{-1}J}.
	\nonumber \\ \label{QQUUQQ_GaussInt}
\eea
The determinant of $M$ can be computed explicitly from \Eq{M}, and is given by
\begin{eqnarray}
	\det M 
	&=&4e^{2i\Delta\theta}\left[ -\sin^2\prn{\Delta\theta} +  \sin^2\prn{2\varphi_a+\Delta\theta}\tanh^2 r_a + \sin^2\prn{2\varphi_b-\Delta\theta}\tanh^2 r_b \right. \nonumber \\
	&&\quad\,\, \left. -2\sin\prn{2\varphi_a}\sin\prn{2\varphi_b}\tanh r_a\tanh r_b - \sin^2\prn{2\varphi_a - 2\varphi_b + \Delta\theta}\tanh^2 r_a\tanh^2 r_b    \right]
	\nonumber \\
&\equiv&	f_M(a,b),
	 \label{detM}
\end{eqnarray}
where $\Delta\theta\equiv\theta_a - \theta_b$. This expression defines the function $f_M(a,b)$, which satisfies $f_M(b,a) = f_M^{*}(a,b)$.

Since the four first entries of $J$ vanish, see \Eq{eq:J:def}, the $8\times 8$ lower right block of $M^{-1}$ is sufficient to compute $J^\mathrm{T}M^{-1}J$. We therefore focus on that  $8\times 8$  block, \ie on the matrix $\tilde M$ defined as $\tilde M_{ij} = \prn{M^{-1}}_{i+4, j+4}$, with $i,j=1\cdots8$. It can be computed explicitly from \Eq{M}, and its expression involves the four functions $d_1,\,d_2,\,d_3,\,d_4$, defined as
\bea
	d_1(a,b)&=&
	e^{2i\Delta\theta}\left\lbrace
	f\prn{2\varphi_a+\Delta\theta}\tanh^2r_a + f\prn{2\varphi_b-\Delta\theta}\tanh^2r_b-f\prn{\Delta\theta}
	\right. \nonumber \\
	&&\left.
	-2\left[f\prn{\varphi_a+\varphi_b}-f\prn{\varphi_b-\varphi_a}\right]\tanh r_a\tanh r_b 
	\right. \nonumber \\ &&\left.
	- f\prn{ 2\varphi_b -2\varphi_a - \Delta\theta}\tanh^2 r_a\tanh^2 r_b \right\rbrace, \label{eq:d1:def} \\
	d_2(a,b)&=&
	 4i e^{2i\Delta\theta} \left[ \sin\prn{2\varphi_a}\tanh r_a 
	 - \sin\prn{2\varphi_b-2\Delta\theta}\tanh r_b \right. \nonumber \\
	 &&\left.- \sin\prn{4\varphi_a - 2\varphi_b + 2\Delta\theta}\tanh^2r_a\tanh r_b 
	+\sin\prn{2\varphi_a}\tanh r_a\tanh^2 r_b    \right],\quad \\
	d_3(a,b)
	&=& \frac{-4i e^{2i\Delta\theta} \com{\sin\Delta\theta + \sin\prn{2\varphi_a - 2\varphi_b+\Delta\theta} \tanh r_a \tanh r_b}}{\cosh r_a\cosh r_b} ,\\
	d_4(a,b)
	&=& \frac{4i e^{2i\Delta\theta} \com{\sin\prn{2\varphi_a+\Delta\theta}\tanh r_a - \sin\prn{2\varphi_b-\Delta\theta}\tanh r_b}}{\cosh r_a\cosh r_b},
	\label{eq:d4:def}
\eea
where $f(\theta) = 1-\cos 2\theta + 2i\sin 2\theta.$\footnote{In practice, the following relations satisfied by the function $f$ turn out to be useful:
\bea
	f(\theta) = 2\sin\theta\prn{\sin \theta + 2i\cos\theta}
	= \frac{1}{2}\prn{e^{2i\theta} - 3e^{-2i\theta}  } + 1
	= 2\com{1 - f\prn{\frac{\theta}{2} - \frac{\pi}{4}}}\sin \theta,
\eea
and
\bea
	f(-\theta) = f^*(\theta)
	,\quad
	f(\theta \pm \pi) = f(\theta).
\eea
}
The matrix $\tilde M$ can be expressed as
\begin{eqnarray}
\label{eq:Mtilde:D}
	\tilde M=\com{
	\begin{array}{cccccccc}
		\frac{1}{2} & \frac{i}{2} & 0 & 0 & 0 & 0 & 0 & 0
		\vspace{1mm}\\
		 \frac{i}{2} & D_1 & 0 & D_2 & 0 & D_3 & 0 & D_4
		\vspace{1mm}\\
		 0 & 0 & \frac{1}{2} & \frac{i}{2} & 0 & 0 & 0 & 0
		\vspace{1mm}\\
		 0 & D_2 & \frac{i}{2} & D_1 & 0 & D_4 & 0 & D_3
		\vspace{1mm}\\
		 0 & 0 & 0 & 0 & \frac{1}{2} & -\frac{i}{2} & 0 & 0
		\vspace{1mm}\\
		 0 & D_3 & 0 & D_4 & -\frac{i}{2} & \bar D_1 & 0 & \bar D_2
		\vspace{1mm}\\
		 0 & 0 & 0 & 0 & 0 & 0 & \frac{1}{2} & -\frac{i}{2}
		\vspace{1mm}\\
		 0 & D_4 & 0 & D_3 & 0 & \bar D_2 & -\frac{i}{2} & \bar D_1
	\end{array}
	},
\end{eqnarray}
where
\bea
\label{eq:Di:def}
	D_i &=& \frac{d_i(a,b)}{f_M(a,b)} \quad \mathrm{for} \quad i = 1,2,3,4\\
	\bar D_i &=& \com{\frac{d_i(b,a)}{f_M(b,a)}}^* = \frac{d_i^*(b,a)}{f_M(a,b)}
	\quad \mathrm{for}  \quad i = 1,2.
\label{eq:Dibar:def}
\eea
One can see that the bars denote the operation of taking the complex conjugate and flipping ``$a$'' and ``$b$''.

The calculation of $J^\mathrm{T}M^{-1}J$ can then be performed as follows. From \Eq{eq:J:def}, $J$ can be written as $J^{\mathrm{T}}=[0,0,0,0,(G\cdot X)^{\mathrm{T}}]$, with
\bea
	G=\com{
	\begin{array}{cccc}
		-1 & 0 & 0 & 0
		\vspace{1mm}\\
		 i & 0 & 0 & 0
		\vspace{1mm}\\
		 0 & -1 & 0 & 0
		\vspace{1mm}\\
		 0 & i & 0 & 0
		\vspace{1mm}\\
		 0 & 0 & -1 & 0
		\vspace{1mm}\\
		 0 & 0 & -i & 0
		\vspace{1mm}\\
		 0 & 0 & 0 & -1
		\vspace{1mm}\\
		 0 & 0 & 0 & -i
	\end{array}
	}
	,\quad
	X = \com{
	\begin{array}{c}
		\bar Q_1
		\vspace{1mm}\\
		\bar Q_2
		\vspace{1mm}\\
		\tilde Q_1
		\vspace{1mm}\\
		 \tilde Q_2
	\end{array}
	}.
\label{eq:G:X}
\eea
Then $J^\mathrm{T}M^{-1}J = X^\mathrm{T}G^\mathrm{T} \tilde M GX=X^\mathrm{T}\mathcal MX$, where $\mathcal M\equiv G^\mathrm{T} \tilde M G$.
$\mathcal M$ can be calculated from \Eqs{eq:Mtilde:D} and~\eqref{eq:G:X}, and one obtains
\bea
 \hspace{-6mm}
	\mathcal M=\com{
	\begin{array}{cccc}
		\frac{3}{2}-\tilde M_{22} & -\tilde M_{24} & \tilde M_{26} & \tilde M_{28} 
		\vspace{1mm}\\
		 -\tilde M_{42} & \frac{3}{2}-\tilde M_{44} & \tilde M_{46} & \tilde M_{48} 
		\vspace{1mm}\\
		 \tilde M_{62} & \tilde M_{64} & \frac{3}{2}-\tilde M_{66} & -\tilde M_{68}
		\vspace{1mm}\\
		\tilde M_{82} & \tilde M_{84} & -\tilde M_{86} & \frac{3}{2}-\tilde M_{88} 
	\end{array}
	}
	=\com{
	\begin{array}{cccc}
		\frac{3}{2}-D_1 & -D_2 & D_3 & D_4 
		\vspace{1mm}\\
		 -D_2 & \frac{3}{2}-D_1 & D_4 & D_3 
		\vspace{1mm}\\
		 D_3 & D_4 & \frac{3}{2}-\bar D_1 & -\bar D_2
		\vspace{1mm}\\
		 D_4 & D_3 & -\bar D_2  & \frac{3}{2}-\bar D_1
	\end{array}
	}.
\eea
Combining the above results together, one obtains
\bea
	\Braket{\tilde Q_1, \tilde Q_2|\hat U(t_a) \hat U^\dagger(t_b)| \bar Q_1, \bar Q_2} =
	\frac{\exp\left(\frac{1}{2}  X^\mathrm{T}\tilde{\mathcal M}X      \right)    }
	{\pi\cosh r_a \cosh r_b\sqrt{\det M}},
	\label{QQUUQQ}
\eea
where
\begin{eqnarray}
	\tilde{\mathcal M} = \com{
	\begin{array}{cccc}
		\frac{1}{2}-D_1 & -D_2 & D_3 & D_4 
		\vspace{1mm}\\
		 -D_2 & \frac{1}{2}-D_1 & D_4 & D_3 
		\vspace{1mm}\\
		 D_3 & D_4 & \frac{1}{2}-\bar D_1 & -\bar D_2 
		\vspace{1mm}\\
		 D_4 & D_3 & -\bar D_2  & \frac{1}{2}-\bar D_1 
	\end{array}
	}.
\end{eqnarray}
This is the result we use in the main text to derive \Eq{lambda_expression}.
\section{Gaussian integral over the quadrants}
\label{app:Gaussian_int}
In this appendix, we consider the following integral
\bea
	I\equiv\int_0^\infty\dd x\int_0^\infty\dd y\,e^{-ax^2-2bxy-cy^2}\, ,
\eea
where $a,b,c\in \mathbb C$. Our goal is to derive a closed-form expression for $I$, and to carefully study the conditions under which that expression is valid. The result is used to derive \Eq{Eab_Bell_analytic}, a formula for the temporal correlation function in the limit of infinite $\ell$.

First, let us introduce the integral
\bea
\label{eq:app:Gaussia:int:j:def}
	J\left(\xi,\beta\right)\equiv\int_0^\infty\dd x\, e^{-\xi x^2}\erf\prn{\beta x}\, ,
\eea
where $\xi, \beta \in \mathbb C$, and where the error function is defined as $\mathrm{erf}(z) = \frac{2}{\sqrt\pi}\int_0^ze^{-t^2}\dd t$. Here, the integration variable $t$ follows a straight line between $0$ and $z$ in the complex plane. The error function is an odd function, \ie $\mathrm{erf}(-z) = -\mathrm{erf}(z)$. By replacing the error function by its definition, and upon changing the order of integration (the conditions for absolute integrability, which ensure the validity of Fubini's theorem, are discussed at the end of this appendix), $J$ can be expressed as
\bea
	J\left(\xi,\beta\right)&=&\frac{2}{\sqrt \pi}\int_0^\infty\dd x\,e^{-\xi x^2}\int_0^{\beta x}\dd t\,e^{-t^2} \nonumber\\
	&=&\frac{2}{\sqrt \pi}\int_0^\infty\dd x\,e^{-\xi x^2} \beta x \int_0^{1}\dd u\,e^{-\beta^2x^2u^2} \nonumber\\
	&=&\frac{2\beta}{\sqrt \pi}\int_0^{1}\dd u\,\int_0^\infty\dd x\,x\,e^{-\prn{\xi+\beta^2u^2} x^2} \nonumber\\
	&=&\frac{2\beta}{\sqrt \pi}\int_0^{1}\dd u\,\com{ \frac{e^{-\prn{\xi+\beta^2u^2} x^2}}{-2\prn{\xi+\beta^2u^2}} }_0^\infty\, ,
\label{eq:J:1}
\eea
where in the second line, we have performed the change of integration variable $t=\beta xu$.
In order for this integral to converge, one must assume $\Re\mathrm e(\xi)>0$ and $\Re\mathrm e(\xi+\beta^2)>0$. Then, one can proceed as
\bea
	J\left(\xi,\beta\right)&=&\frac{\beta}{\sqrt \pi}\int_0^{1} \frac{\dd u}{\xi+\beta^2u^2} \nonumber\\
	&=&\frac{1}{\sqrt \pi \sqrt\xi}\int_0^{\beta/\sqrt\xi} \frac{\dd v}{1+v^2}\, ,
\eea
where we have performed the change of integration variable $v= \beta u/\sqrt \xi$.
In this last expression, let us note that the result of the complex integral depends on the path followed in the complex plane, since the integrand has poles at $v = \pm i$. In the present case however, as mentioned above, the path is a straight line, which leaves no ambiguity. 
This integral can then be expressed in terms of the arctangent function,
\bea
\label{eq:J:arctan}
	J\left(\xi,\beta\right)&=&\frac{1}{\sqrt \pi \sqrt\xi}\,\mathrm{arc}\tan\prn{\frac{\beta}{\sqrt\xi}}\, .
\eea
Here, the range of the real part of the arctangent is restricted to $[-\pi/2, \pi/2]$ as usual.

Let us now introduce a second integral, $J_\uc$, defined similarly to $J$ but where the error function is replaced with the complementary error function
\bea
	J_\uc\left(\xi,\beta\right)\equiv\int_0^\infty\dd x\, e^{-\xi x^2}\erfc\prn{\beta x}\, .
\eea
The complementary error function is related to the error function by $\erfc(z)=1-\erf(z)$. By making use of \Eq{eq:J:arctan} and of the relation [see Eq.~(4.24.17) of \Refa{2011ConPh..52..497T}]
\bea
	\mathrm{arc}\tan(z)+\mathrm{arc}\tan\prn{\frac{1}{z}}=
	\begin{cases}
		\pi / 2 & \mathrm{for}\quad \Re\mathrm e (z)>0\\
		-\pi / 2 & \mathrm{for}\quad \Re\mathrm e (z)<0
	\end{cases}\nonumber ,
\eea
one has
\bea
	J_\uc\left(\xi,\beta\right)
	&=&\int_0^\infty\dd x\, e^{-\xi x^2}\com{1-\erf\prn{\beta x}} \\
	&=&\frac{\sqrt \pi}{2\sqrt\xi}-J\left(\xi,\beta\right) 
	= \frac{\sqrt \pi}{2\sqrt\xi}-\frac{1}{\sqrt \pi\sqrt\xi}\,\mathrm{arc}\tan\prn{\frac{\beta}{\sqrt\xi}}
	\label{Gauss_erfc_int} \\
	&=&\begin{cases}
	\displaystyle\frac{1}{\sqrt \pi \sqrt \xi}\,\mathrm{arc}\tan\prn{\frac{\sqrt\xi}{\beta}}
	& \displaystyle\mathrm{for}\quad \Re\mathrm e\prn{\frac{\beta}{\sqrt\xi}}>0 
	\vspace{1mm}\\
	\displaystyle\frac{\sqrt \pi}{\sqrt\xi}+\frac{1}{\sqrt \pi \sqrt \xi}\,\mathrm{arc}\tan\prn{\frac{\sqrt\xi}
	{\beta}}
	& \displaystyle\mathrm{for}\quad \Re\mathrm e\prn{\frac{\beta}{\sqrt\xi}}<0
	\end{cases} \label{Gauss_erfc_int_branch}
	.
\eea

We now apply these results to the calculation of the integral $I$. By noticing that
\bea
	-ax^2-2bxy-cy^2 = -c\prn{y+\frac{bx}{c}}^2+\prn{\frac{b^2}{c}-a}x^2,
\eea
one obtains
\bea
	I&=&\int_0^\infty\dd x\,e^{-\prn{a-\frac{b^2}{c}}x^2}\int_0^\infty\dd y\,e^{-c\prn{y+\frac{bx}{c}}^2} \nonumber\\
	&=&\int_0^\infty\dd x\,e^{-\prn{a-\frac{b^2}{c}}x^2}\frac{1}{\sqrt c}\int_{bx/\sqrt c}^{\sqrt c\prn{\infty+bx/c}}\dd z\,\ee^{-z^2} ,
\eea
where in the last expression, we have performed the change of integration variable $z=\sqrt c\prn{y+\frac{bx}{c}}$. Assuming that $\Re\mathrm e(c)>0$, one can write the integral over $z$ in terms of the complementary error function,\footnote{Note that in the limit $r\rightarrow\infty$, one has [see Eq.~(7.12.1) of \Refa{2011ConPh..52..497T}]
\bea
	\erf\prn{r e^{i\theta}}
	\underset{r\to\infty}{\longrightarrow}
	\begin{cases}
		1 & \mathrm{for}\quad -\pi/4<\theta<\pi/4\\
		\mathrm{divergent} & \mathrm{otherwise}
	\end{cases}\nonumber .
\eea
}
\begin{eqnarray}
	I&=&\frac{\sqrt\pi}{2\sqrt c}\int_0^\infty\dd x\,e^{-\prn{a-\frac{b^2}{c}}x^2}\erfc\prn{\frac{b}{\sqrt c}x}
	=\frac{\sqrt\pi}{2\sqrt c} J_\uc\left(a-\frac{b^2}{c},\frac{b}{\sqrt c}\right).
\end{eqnarray}
According to the conditions given below \Eq{eq:J:1}, this expression is well defined if $\Re\mathrm e(a)>0$ and $\Re\mathrm e\prn{a-b^2/c}>0$. Making use of \Eq{Gauss_erfc_int}, one obtains
\begin{eqnarray}
	I=\frac{1}{2\sqrt c \sqrt{a-\frac{b^2}{c}}}\com{\frac{\pi}{2}-\mathrm{arc}\tan\prn{\frac{b}{\sqrt c\sqrt{a-\frac{b^2}{c}}}}}.
	\label{Gauss_int_preconclusion}
\end{eqnarray}
Note that $\sqrt c \sqrt{a-b^2/c} = \sqrt{ac - b^2}$ without ambiguity on the sign (in general, the branch cut of the complex square root function leaves the sign of $\sqrt{z}$ ambiguous) under the assumptions $\Rea (c)>0, \Rea \prn{a-b^2/c}>0$.

In summary, we have showed that
\begin{eqnarray}
	\int_0^\infty\dd x\int_0^\infty\dd y\,e^{-ax^2-2bxy-cy^2}=\frac{1}{2 \sqrt{ac-b^2}}\com{\frac{\pi}{2}-\mathrm{arc}\tan\prn{\frac{b}{ \sqrt{ac-b^2}}}}
	\label{Gauss_int_1}
\end{eqnarray}
under the conditions
\begin{eqnarray}
	\Re\mathrm e(a)>0 ,\quad
	\Re\mathrm e(c)>0 ,\quad
	\Re\mathrm e\prn{a-\frac{b^2}{c}}>0.
	\label{Gauss_int_cond}
\end{eqnarray}
Let us note that the integral over other quadrants can be derived following the same lines. For instance, one has
\begin{eqnarray}
	\int_{-\infty}^0\dd x\int_0^\infty\dd y\,e^{-ax^2-2bxy-cy^2}
	&=&\int_0^\infty\dd x\int_0^{\infty}\dd y\,e^{-ax^2+2bxy-cy^2} \nonumber\\
	&=&\frac{1}{2\sqrt{ac-b^2}}\com{\frac{\pi}{2}+\mathrm{arc}\tan\prn{\frac{b}{\sqrt{ac-b^2}}}}
	\label{Gauss_int_2}
\end{eqnarray}
under the same conditions~\eqref{Gauss_int_cond}. These expressions could be further simplified, as in \Eq{Gauss_erfc_int_branch}, but one would then have to consider two branches, and we do not display the resulting formulas since they are not particularly insightful. The integrals over the two remaining quadrants can be readily derived from \Eqs{Gauss_int_1} and~\eqref{Gauss_int_2} by exchanging the integration variables $x$ and $y$, \ie by swapping $a$ and $b$ in the formulas. In summary, for the integrals over the four quadrants to be well defined, the condition~\eqref{Gauss_int_cond} must be satisfied {\it per se} and also after exchanging $a$ and $c$, which leads to
\bea
	\Re\mathrm e(a)>0 ,\quad
	\Re\mathrm e(c)>0 ,\quad
	\Re\mathrm e\prn{a-\frac{b^2}{c}}>0,\quad
	\Re\mathrm e\prn{c-\frac{b^2}{a}}>0 .
	\label{Gauss_int_cond_refined}
\eea

\par $ $\\
Let us finally discuss the convergence conditions for the integrals studied in this appendix. According to Fubini's theorem, a double integral can be evaluated by means of an iterated integral if the integrand is absolutely integrable. If one were to evaluate the Gaussian integral over the full two-dimensional plane, the condition for absolute convergence would be
\begin{eqnarray}
    \Rea (a)>0 ,\quad
	\Rea (c)>0 ,\quad
	\Rea \prn{a}\Re\mathrm e\prn{c}-\com{\Rea \prn{b}}^2>0.
	\label{abs_convergence_cond}
\end{eqnarray}
One should note that \Eq{Gauss_int_cond_refined} is always true if \Eq{abs_convergence_cond} is satisfied, for the following reason. The conditions on $\Rea (a)$ and $\Rea (c)$ being the same, one needs to focus on the third and fourth conditions in \Eqs{Gauss_int_cond_refined}, and on the third condition in \Eq{abs_convergence_cond}. By expanding $b$ and $c$ into their real and imaginary parts, one has
\bea
\label{eq:minimize:Re(a-b2Overc)}
\Rea \prn{a-\frac{b^2}{c}} = \Rea(a) - \frac{\Rea^2(b) \Rea(c) - \Ima^2(b) \Rea(c) - 2 \Rea(b) \Ima(b) \Ima(c)}{\Rea^2(c)+\Ima^2(c)}\, .
\eea
Let us view this expression as a function of $\Ima(b)$. Its derivative vanishes when $\Ima(b) = - \Rea(b) \Ima(c) / \Rea(c)$, and at that point, the second derivative reads $2 \Rea(c)/[\Rea^2(c)+\Ima^2(c)]$. Under the condition $\Rea(c)>0$, which is contained in both \Eqs{Gauss_int_cond} and~\eqref{abs_convergence_cond}, the second derivative is thus positive, hence $\Ima(b) = - \Rea(b) \Ima(c) / \Rea(c)$ is a global minimum. Evaluating \Eq{eq:minimize:Re(a-b2Overc)} at that point, one thus obtains
\bea
\label{eq:cond:ineq}
\Rea \prn{a-\frac{b^2}{c}} > \Rea(a)-\frac{\Rea^2(b)}{\Rea(c)}\quad\quad\mathrm{if}\quad\quad\Rea(c)>0\, .
\eea
Since $\Rea(c)>0$ in \Eq{abs_convergence_cond}, the third condition in \Eq{abs_convergence_cond} is equivalent to the requirement that the right-hand side of \Eq{eq:cond:ineq} is positive, and this implies the validity of the third condition in \Eq{Gauss_int_cond_refined}. Since the third condition in \Eq{abs_convergence_cond} is symmetric in $a$ and $b$, this also implies the validity of the fourth condition in \Eq{Gauss_int_cond_refined}, which finishes to prove that \Eq{abs_convergence_cond} implies \Eq{Gauss_int_cond_refined}.

The cases where \Eq{Gauss_int_cond_refined} is valid while \Eq{abs_convergence_cond} is not are beyond the scope of Fubini's theorem and there, the correctness of the calculation performed in this appendix is a priori nontrivial. However, when this happens, we have checked with a direct numerical integration that our formulae are still valid. 
While \Eq{abs_convergence_cond} is a sufficient condition for making use of Fubini's theorem, it is not always necessary, and our results thus suggest that \Eq{Gauss_int_cond_refined} is a necessary, and possibly sufficient, condition.
In every physical situation we have looked at, we have checked that the conditions~\eqref{Gauss_int_cond_refined} are satisfied, which ensures that finite results are obtained.
\section{Derivation of the small-$\ell$ expansion formula}
\label{app:small_ell}
In this appendix, we consider the small-$\ell$ limit of integrals of the form
\bea
	I\equiv \sum_{n = -\infty}^\infty \sum_{m = -\infty}^\infty  (-1)^{n+m}
	\int_{n\ell}^{(n+1)\ell}\dd x\int_{m\ell}^{(m+1)\ell}\dd y
	\,e^{-ax^2-2bxy-cy^2}\, ,
\eea
where $a,b,c\in \mathbb C$. Let us first perform the change of integration variables $x' = (x -n\ell)/\ell$ and $y' = (y -n\ell)/\ell$, which allows us to rewrite $I$ as
\bea
\label{eq:app:low:l:change:integration:variable:1}
	I&=& \ell^2 \sum_{n = -\infty}^\infty \sum_{m = -\infty}^\infty  (-1)^{n+m}
	\int_{0}^{1}\dd x\int_{0}^{1}\dd y
	\,e^{-a(x+n)^2\ell^2-2b(x+n)(y+m)\ell^2 -c(y+m)^2\ell^2}\\
\label{eq:app:low:l:change:integration:variable:2}
	&=& \ell^2\int_{0}^{1}\dd x\int_{0}^{1}\dd y \,e^{-(ax^2+2bxy+cy^2)\ell^2}
	\nonumber \\
	&&\sum_{n = -\infty}^\infty  (-1)^{n} e^{-2(ax+by)n\ell^2-an^2\ell^2}
	\sum_{m = -\infty}^\infty  (-1)^{m} e^{-2(bx + bn + cy)m\ell^2 - c m^2\ell^2}\\
\label{eq:app:low:l:change:integration:variable:3}
	&=& \ell^2\int_{0}^{1}\dd x\int_{0}^{1}\dd y \,e^{-(ax^2+2bxy+cy^2)\ell^2}
	\nonumber \\
	&&
	\sum_{n = -\infty}^\infty  (-1)^{n} e^{-2(ax+by)n\ell^2-an^2\ell^2}
	\vartheta_4\left[ i(bx + bn + cy)\ell^2, e^{-c\ell^2}\right]\, .
\eea
Here, in \Eq{eq:app:low:l:change:integration:variable:2}, we have exchanged the order by which we integrate over $x$ and $y$ and sum over $n$ and $m$, which is possible since all integrals and sums are absolutely convergent under the conditions detailed in \App{app:Gaussian_int}, and in \Eq{eq:app:low:l:change:integration:variable:3}, we have recast the sum over $m$ in terms of an elliptic theta function.\footnote{Hereafter we make use of the two elliptic theta functions
\begin{eqnarray}
\label{eq:vartheta:4:def}
	\vartheta_4(z,q)&\equiv&\sum_{n = -\infty}^\infty (-1)^n q^{n^2}e^{2inz}\, , \\
\label{eq:vartheta:2:def}
	\vartheta_2(z,q)&\equiv&q^{1/4}e^{iz}\sum_{n = -\infty}^\infty q^{n(n+1)}e^{2inz} \, ,
\end{eqnarray}
where $z,q\in\mathbb C$ and $\abs{q}<1$.}
Using the Jacobi identity [see Eq.~(20.7.33) of \Refa{2011ConPh..52..497T}]
\bea
	\vartheta_4\prn{z, e^{i\pi\tau}} = (-i\tau)^{-1/2}e^{-\frac{iz^2}{\pi\tau}}\vartheta_2\prn{-\frac{z}{\tau}, e^{-\frac{i\pi}{\tau}}},
\eea
one can write
\bea
	\vartheta_4\left[ i(bx + bn + cy)\ell^2, e^{-c\ell^2}\right] = \frac{\sqrt\pi}{\ell\sqrt{c}}
	\,e^{(bx + bn + cy)^2\frac{\ell^2}{c}} \, \vartheta_2\left[-\frac{\pi(bx + bn + cy)}{c}, e^{-\frac{\pi^2}{c\ell^2}}\right]\, .
\eea
This allows us to obtain an expression in which the first argument of the elliptic theta function is independent of $\ell$, and the second argument tends to $0$ when $\ell$ tends to $0$. In \Eq{eq:vartheta:2:def}, when $\vert q \vert \ll 1$ , the two dominants terms are the ones with $n=0$ and $n=-1$, which gives rise to 
\bea
\label{eq:vartheta2:smallq:expansion}
	\vartheta_2(z,q)\simeq2q^{1/4}\cos z \qquad \mathrm{for} \qquad \abs{q}\ll1.
\eea
Combining these results, one obtains
\bea
\label{eq:app:low:l:elliptic:resum:1}
	I&\simeq& \frac{2\sqrt\pi \ell}{\sqrt{c}}e^{-\frac{\pi^2}{4c\ell^2}} \int_{0}^{1}\dd x \, e^{-\left(a-\frac{b^2}{c}\right)x^2\ell^2}
	\sum_{n = -\infty}^\infty  (-1)^{n} e^{-\left(a-\frac{b^2}{c}\right)(n^2+2xn)\ell^2}
	\nonumber \\
	&&\int_{0}^{1}\dd y \, \cos\left[\pi y + \frac{\pi b(x+n)}{c}\right]\\
\label{eq:app:low:l:elliptic:resum:2}
	&=& \frac{-4 \ell}{\sqrt \pi \sqrt{c}}e^{-\frac{\pi^2}{4c\ell^2}} \int_{0}^{1}\dd x \, e^{-\left(a-\frac{b^2}{c}\right)x^2\ell^2}
	\sum_{n = -\infty}^\infty  (-1)^{n} e^{-\left(a-\frac{b^2}{c}\right)(n^2+2xn)\ell^2}
	\sin\left[\frac{\pi b(x+n)}{c}\right] \qquad\, \\
\label{eq:app:low:l:elliptic:resum:3}
	&=& \frac{2i \ell}{\sqrt \pi \sqrt{c}}e^{-\frac{\pi^2}{4c\ell^2}} \int_{0}^{1}\dd x \, e^{-\left(a-\frac{b^2}{c}\right)x^2\ell^2}
	\prn{J_+ - J_-}\, .
\eea
Here, in \Eq{eq:app:low:l:elliptic:resum:2}, we have performed the integral over $y$, and in \Eq{eq:app:low:l:elliptic:resum:3}, we have expanded the $\sin$ function in terms of exponentials, and introduced
\bea
	J_\pm &\equiv& \sum_{n = -\infty}^\infty  (-1)^{n} e^{-\left(a-\frac{b^2}{c}\right)(n^2+2xn)\ell^2 \pm i\pi b\frac{x+n}{c} }\\
	&=& e^{\pm i\pi bx/c} 
	\,\vartheta_4\left[i\prn{a-\frac{b^2}{c}}x\ell^2 \pm \frac{\pi b}{2 c}, e^{-\left(a-\frac{b^2}{c}\right)\ell^2}\right] 
\eea
where \Eq{eq:vartheta:4:def} has been used to express the sum over $n$ as an elliptic theta function.

If one further assumes that $a-b^2/c$ has a positive real part, one can make use of the expansion formula\footnote{We make use of Eq.~(2.7.33) of \Refa{2011ConPh..52..497T}, namely
\bea
\left(-i\tau\right)^{1/2} \vartheta_4(z,q) = \ee^{i\frac{\tau' z^2}{\pi}} \vartheta_2\left(z\tau',q'\right)\, ,
\eea
where $q$ and $\tau$ are related through $q=\ee^{i\pi\tau}$, and a similar relation for $q'$ and $\tau'$, and where $\tau'=-1/\tau$. Denoting $q=\ee^{-\alpha}$, one has $q'=\ee^{-\pi^2/\alpha}$. So when $\vert \alpha \vert \ll 1$, $\vert q' \vert \ll 1$ if $\Rea(\alpha)>0$. In this limit, one can expand the $\vartheta_2$ function according to \Eq{eq:vartheta2:smallq:expansion}, and this gives rise to \Eq{eq:expansion:theta4:alpha}.
}
\bea
	\vartheta_4(z,e^{-\alpha})\simeq\frac{2\sqrt\pi}{\sqrt{\alpha}}e^{-\frac{\pi^2+4z^2}{4\alpha}}\cosh\prn{\frac{\pi z}{\alpha}}
	\qquad \mathrm{for} \qquad \Rea(\alpha)>0, \ \ \ \vert \alpha \vert \ll 1 ,
\label{eq:expansion:theta4:alpha}
\eea
and rewrite $J_\pm$ as
\bea
	J_\pm \simeq \frac{2\sqrt\pi}{\ell\sqrt{a-\frac{b^2}{c}}}
	\exp\com{-\frac{\pi^2\left(1+\frac{b^2}{c^2}\right)}{4\left(a-\frac{b^2}{c}\right)\ell^2} + \prn{a-\frac{b^2}{c}}x^2\ell^2} C_\pm\, ,
\eea
where
\bea
	C_\pm \equiv \cosh\left[i\pi x \pm \frac{\pi^2 b}{2(ac - b^2)\ell^2}\right]\, .
\eea
Combining the above results, one obtains
\bea
	I\simeq \frac{4i}{\sqrt c\sqrt{a-\frac{b^2}{c}}} \exp\com{-\frac{\pi^2(a+c)}{4(ac-b^2)\ell^2}}
	\int_0^1 \dd x\,\prn{C_+ - C_-}\, ,
\eea
where the integral over $x$ can be performed analytically and gives rise to
\bea
	\int_0^1 \dd x\,C_\pm = \pm\frac{2i}{\pi}\sinh\left[\frac{\pi^2 b}{2(ac - b^2)\ell^2}\right] .
\eea
Here again, as we mentioned below \Eq{Gauss_int_preconclusion}, $\sqrt c \sqrt{a-b^2/c} = \sqrt{ac - b^2}$ holds under the assumptions $\Rea (c)>0, \Rea \prn{a-b^2/c}>0$, which we need for the integral to be convergent.
One finally has
\bea
	I&\simeq& \frac{-16}{ \pi \sqrt{ac - b^2}} \exp\com{-\frac{\pi^2(a+c)}{4(ac-b^2)\ell^2}}
	\sinh\left[\frac{\pi^2 b}{2(ac - b^2)\ell^2}\right] \\
	&=& \frac{8}{ \pi \sqrt{ac - b^2}}\prn{e^{p_+} - e^{p_-}}\, ,
\label{eq:app:low:l:result}
\eea
where
\bea
\label{eq:app:low:l:p:pm}
	p_\pm\equiv -\frac{\pi^2(a+c\pm 2b)}{4(ac-b^2)\ell^2}\, .
\eea
These expressions are used to derive \Eq{E12_small_ell}.
\bibliographystyle{JHEP}
\bibliography{TemporalBell}
\end{document}

%% file: newcommands.tex
\newcommand{\ie}{\textsl{i.e.~}}

\newcommand{\eg}{\textsl{e.g.~}}

\DeclareMathOperator{\erf}{erf}

\newcommand{\dd}{\mathrm{d}}
\newcommand{\ee}{e}

\newcommand{\umin}{\mathrm{min}}
\newcommand{\umax}{\mathrm{max}}

\newcommand{\uc}{\mathrm{c}}

\newcommand{\Rea}{\Re \mathrm{e}\,}
\newcommand{\Ima}{\Im \mathrm{m}\,}

\newcommand{\beq}{\begin{equation}}
\newcommand{\eeq}{\end{equation}}
\newcommand{\bea}{\begin{eqnarray}}
\newcommand{\eea}{\end{eqnarray}}

\newlength{\wsingfig}
\newlength{\wdblefig}
\newlength{\wquadfig}
\newlength{\wtriplefig}
\newcommand{\Eq}[1]{Eq.~(\ref{#1})}
\newcommand{\Eqs}[1]{Eqs.~(\ref{#1})}
\newcommand{\Fig}[1]{Fig.~{\ref{#1}}}

\newcommand{\Refa}[1]{Ref.~{\cite{#1}}}
\newcommand{\Refs}[1]{Refs.~{\cite{#1}}}
\newcommand{\Sec}[1]{Sec.~\ref{#1}}
\newcommand{\Secs}[1]{Secs.~\ref{#1}}
\newcommand{\App}[1]{Appendix~\ref{#1}}